\documentclass[aps,pre,longbibliography,twocolumn,floatfix,superscriptaddress,nofootinbib]{revtex4-2}
\pdfoutput=1

\usepackage{preamble}
\usepackage{results_diff_sigma}

\def\supplementfilename{sm}

\myexternaldocument{\supplementfilename}

\pdfximage{\supplementfilename.pdf}
\def\numbersupplementpages{\the\pdflastximagepages}

\newif\ifarXiv
\arXivtrue

\begin{document}

\title{Evidence that the AT transition disappears below six dimensions}

\author{Bharadwaj Vedula}
\affiliation{Department of Physics, Indian Institute of Science Education and Research, Bhopal, Madhya Pradesh 462066, India}
\author{M. A. Moore}
\affiliation{Department of Physics and Astronomy, University of Manchester, Manchester M13 9PL, United Kingdom}
\author{Auditya Sharma}
\affiliation{Department of Physics, Indian Institute of Science Education and Research, Bhopal, Madhya Pradesh 462066, India}
\date{\today}

\begin{abstract}
  One of the key predictions of Parisi's broken replica symmetry
  theory of spin glasses is the existence of a phase transition in an
  applied field to a state with broken replica symmetry. This
  transition takes place at the de Almeida-Thouless (AT) line in the
  $h-T$ plane.  We have studied this line in the power-law diluted
  Heisenberg spin glass in which the probability that two spins
  separated by a distance $r$ interact with each other falls as
  $1/r^{2\sigma}$. In the presence of a random vector-field of
  variance $h_r^2$ the phase transition is in the universality class
  of the Ising spin glass in a field. Tuning $\sigma$ is equivalent to
  changing the dimension $d$ of the short-range system, with the
  relation being $d =2/(2\sigma -1)$ for $\sigma < 2/3$.  We have
  found by numerical simulations that
  $h_{\text{AT}}^2 \sim (2/3 -\sigma)$ implying that the AT line does
  not exist below $6$ dimensions and that the Parisi scheme is not
  appropriate for spin glasses in three dimensions.

\end{abstract}

\maketitle

\section{Introduction}
\label{sec:introduction}
\begin{figure}  
  \includegraphics[width=0.48\textwidth]{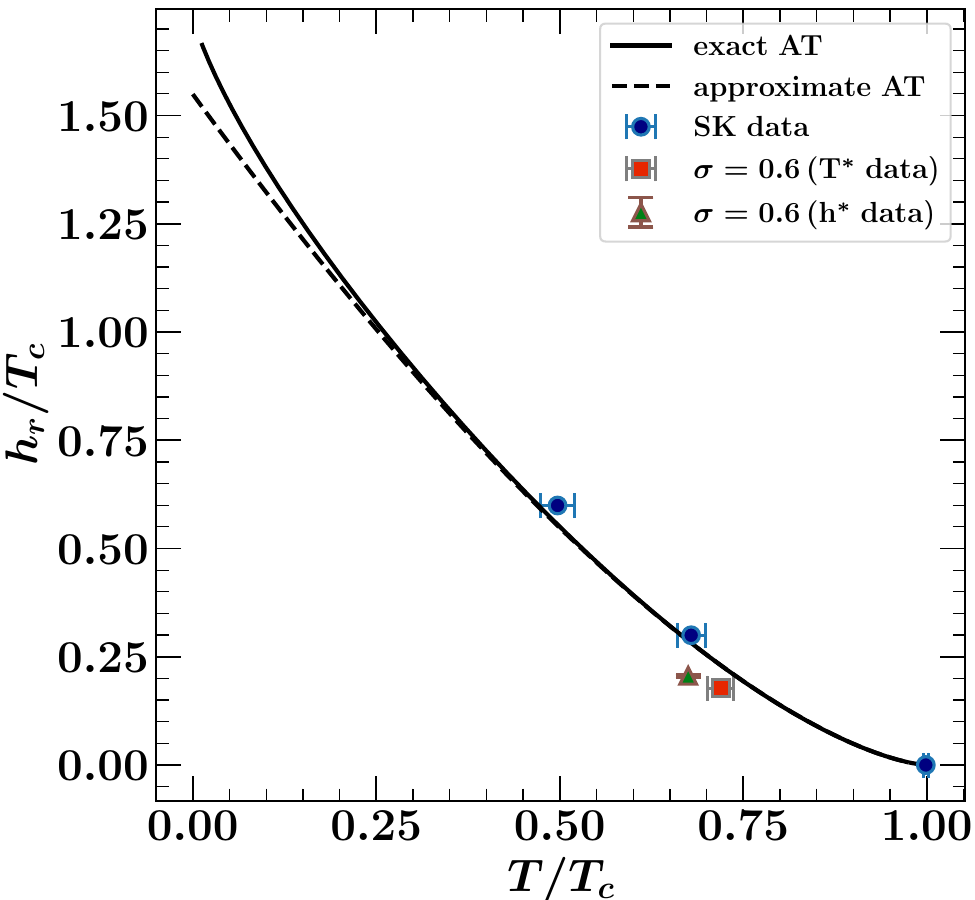}
  \caption{The AT line. The solid line is the exact AT line for the SK
    model, calculated as in Ref. \cite{sharma2010almeida}. The dashed
    line is the approximation to it of Eq.~(\ref{eq:approx_ATMFT})
    with $\zeta=3$. Marked on the diagram by filled circles are the
    results of the simulations on the SK model in
    Ref. \cite{sharma2010almeida}. The red square point, derived from
    varying the temperature $T$ at fixed $h_r$, and the upwards arrow
    point, derived from varying the field at fixed temperature, are
    the result of our simulations at $\sigma = 0.6$, which despite
    corresponding to 10 dimensions, have values of $h_{\text{AT}}$
    suppressed by fluctuations from those which would be estimated
    from the SK model when only adjusting the zero-field transition
    temperature $T_c$.}
  \label{fig:Heisenberg_AT}
\end{figure}

The relevance of the replica symmetry breaking (RSB) scheme of Parisi
\citep{Parisi:79,mezard1987spin} for physical spin glasses in three
dimensions has occasioned doubts from its earliest days
\citep{brayroberts:80}. These doubts have mostly arisen from studies
of the de Almeida-Thouless (AT) line \citep{de1978stability}. This is
the line in the field $h$ and temperature $T$ plane where the
replica symmetric high-temperature phase changes to a phase with broken
replica symmetry (see Fig. \ref{fig:Heisenberg_AT}). The Parisi scheme
has now been rigorously proved to solve the Sherrington-Kirkpatrick
(SK) mean-field model \citep{sherrington:75}, in which all spins
interact with each other. In that model in the presence of a field
$h$, the AT line $h_{\text{AT}}(T)$ for temperatures $T$ close to
$T_c$, the zero-field transition temperature, takes the form
\begin{equation} 
  \left( \frac{h_{\text{AT}}(T)}{T_c} \right)^2 = A(d) \left( 1-\frac{T}{T_c} \right)^\zeta.
  \label{eq:approx_ATMFT}
\end{equation}
The exponent $\zeta=3$ in the SK model and remains at 3 for all $d >
8$.  It takes the value $d/2-1$ when $8 >d > 6$
\citep{Green:83,FisherSomp:85}. For $d < 6$, should the AT line then
still exist, $\zeta = \gamma + \beta$, where the exponent $\gamma$
describes the divergence of the zero-field spin glass susceptibility
$\chi_{\text{SG}}$ as $ T \to T_c$, and $\beta$ describes how the
Edwards-Anderson order parameter $q_{\text{EA}}$ goes to zero in the
same limit \cite{FisherSomp:85}. Both these zero-field exponents have
an expansion in powers of $\epsilon$ where $d =6-\epsilon$
\citep{harrislubchen:76}. Back in 1980 Bray and Roberts
\citep{brayroberts:80} were unable to find a fixed point for the
exponents at the AT line. One possibility which they suggested as an
explanation was that for $d < 6$ there simply was no AT line. However,
the possibility that there was a non-perturbative fixed point could
not be ruled out (but if such exists, it still remains to be
discovered).

Another argument suggested long ago was that of Moore and Bray
\citep{Moore:85}. In $d <6$ the dependence on $\gamma$ and $\beta$ of
the form of the AT line as $\zeta=\gamma +\beta$ indicates that the
applied field $h$ has the scaling dimension of the ordering field of
the spin glass. For $d> 6$ that is not the case, as then
$\gamma +\beta=2$ for all $d > 6$. Usually when the ordering field is
present there is no phase transition. For example, for a ferromagnet
in its ordering field (which is a uniform field) there is no phase
transition as the temperature is lowered. A phase transition only
occurs for vanishing field. The suggestion of Moore and Bray was that
because the applied field had the scaling dimensions of the ordering
field in dimensions $d < 6$ then there would also be no phase
transition in a field and hence no AT line when $d < 6$
\citep{Moore:85}. Even though it is commonplace that a phase
transition is removed in the presence of the ordering field,
alternatives are possible and some were discussed in
Ref. \cite{yeomoore:15}, but no evidence for them was found.

Another argument that $6$ might be the dimension above which RSB
applies comes from RSB calculations of the interface free energy of
the Ising spin glass in the presence of no external field. The
calculations which are valid for $d > 6$ are done for a system of
length $L$ in one direction and $M$ in the other $d-1$ directions. The
interface free energy $\Delta F$ is the change in the free energy when
the boundary conditions in the $L$ direction are switched from
periodic to anti-periodic \cite{aspelmeier:03,aspelmeier:16}. Its
bond-average variance $\Delta F^2$ is found to be of the form
\cite{aspelmeier:16}
\begin{equation}
\Delta F^2 \sim N^{1/3} +L^2 f(L/M),
\label{eq;interface}
\end{equation}
where $N= L M^{d-1}$. The leading term $\sim N^{1/3}$ is of an unusual
form for an interface free energy as it only depends on the volume or
number of spins in the system, and is of this form because interfaces
according to RSB are space filling as their fractal dimension
$d_s= d$. The second term is of the conventional aspect ratio scaling
form which involves the ratio of $L/M$ usually associated with
interfaces whose fractal dimensions $d_s$ are less than $d$ and which
are not space filling \cite{carter:02, hartmann:02}. Using a simple
(and approximate) renormalization group procedure it has been found
that $d_s \to d$ as $d \to 6$ from below \cite{wang:18}. The term
$L^2 f(L/M)$ becomes $L^{2 \theta} f(L/M)$ for $ d < 6$, and then the
first term will not be present. $\theta$ is the interface free energy
exponent \cite{carter:02}.  According to the numerical study of
Boettcher \cite{Boettcher:05} $\theta= 1.1(1)$ in six dimensions,
which suggests it might be exactly $1$ when $d = 6$. But the crucial
point is that for $d > 6$ the first term dominates, but the second
term becomes just as large as the first term right in $d = 6$ for
$L\sim M$. This suggests that $d =6$ is at least an important
dimension for RSB in spin glasses and possibly its lower critical
dimension, the dimension below which full replica symmetry breaking of
the Parisi type is no longer to be found.

Interface free energies are determined by the nature of the
zero-temperature fixed point of the system and its associated
exponents such as $\theta$. These exponents should be distinguished
from those associated with the critical fixed point. The study of Bray
and Roberts \cite{brayroberts:80} was an expansion about the upper
critical dimension of the AT line, which was taken to be 6. The
argument of this paper that there is no RSB for $ d \le 6$, which
if valid implies that the upper and lower critical dimensions for RSB
behavior are both the same and equal to $6$ -- a most unusual
situation!

If the lower critical dimension for the existence of the AT line is
six, then one would expect that the AT line will become closer to the
temperature axis as $d \to 6$. To see whether this is the case
requires determination of the coefficient $A(d)$, but this is very
challenging. In the SK limit for unit length $m$-component vector
spins, $A(d)= 4 m/(m+2)$: For the Heisenberg model studied in this
paper $m=3$. By using an expansion in $1/m$, Moore argued that as
$d \to 6$ from above $A(d) \sim (d-6)$ \citep{moore:12}.  The
numerical studies reported in this paper are consistent with this
possibility. They indeed imply therefore that the AT line is
approaching the temperature axis as $d\to 6$, and hence that there
will not be an AT transition below six dimensions.

The question of whether there is or is not an AT line in physical
dimensions such as $d=3$ has naturally been studied by both experiment
and by simulations. On the experimental side a negative answer was
suggested by the work in Ref. \citep{Nordblad:95}, while a positive
answer was provided in Ref. \citep{Orbach:02}. No consensus is found
in simulations either: for a recent review see
\citep{martinmayor2022numerical}.

Because it is hard to do simulations above 6 dimensions (although
recently an attempt was made to study the AT line in 6 dimensions
\cite{Martin-Mayor:23}), we have done simulations on the
one-dimensional proxy model where systems of large linear extent $L$
can be studied. In Ref. \cite{Martin-Mayor:23} where a six-dimensional
version was directly simulated, $L$ was less than 8, but we can study
values of $L$ up to $65536$. 

We organised the paper into the following sections. In
Sec. \ref{sec:model-hamiltonian} we describe the model we used in
detail. The quantities we studied and their finite size scaling forms
near the AT transition point are given in Sec. \ref{sec:chi-xi}. In
Sec. \ref{fss-analyses} we show the results obtained by performing
finite size scaling analyses on the data for five values of $\sigma$
in the mean-field regime: 0.600, 0.630, 0.640, 0.650, and 0.655. In
Sec. \ref{sec:A_sigma_analysis}, we show our analysis of $A(\sigma)$
versus $\sigma$ which provided us strong evidence that the AT line
disappears below $\sigma_c=2/3$. In an earlier investigation on the XY
model \cite{ XY:23} we had studied it for $\sigma$ values 0.60, 0.70,
0.75 and 0.85, and had observed that because the leading correction to
scaling exponent $\omega$ approaches 0 as $\sigma \to \sigma_c = 2/3$
it would be very challenging to determine whether the AT line goes
away precisely at $\sigma = 2/3$. This means that as $\sigma \to 2/3$
one needs to go to ever larger values of the system size $N$ to
maintain the same level of accuracy. Finally in Sec. \ref{sec:summary}
we summarize our conclusions.

\section{Model Hamiltonian}
\label{sec:model-hamiltonian}

The Hamiltonian of our system is
\begin{equation}
  \mathcal{H}=-\sum\limits_{\langle i,j \rangle}J_{ij} \textbf{S}_i \cdot \textbf{S}_j
  -\sum\limits_{i}\textbf{h}_i\cdot\textbf{S}_i \,,
  \label{eqn:vector_sg_hamiltonian}
\end{equation}
where $\textbf{S}_i$, a unit vector of $m=3$ components, is a spin
sitting on the $i^{\text{th}}$ lattice site ($i=1,2,\ldots,N$). The
$N (\equiv L) $ lattice sites are arranged around a ring of
circumference $N$. So the distance between the
spins at sites $i$ and $j$~\cite{PhysRevB.67.134410}
\begin{equation}
  r_{ij}=\frac{N}{\pi}\sin\left(\frac{\pi}{N}\left| i-j \right| \right),
  \label{eqn:distance}
\end{equation}
is the length of the chord connecting them. The probability of having
a non-zero interaction between a pair of spins $(i,j)$ falls with the
distance $r_{ij}$ between the spins as a power law:
\begin{equation}
  p_{ij}=\frac{r_{ij}^{-2\sigma}}{\sum\limits_{j\neq i}r_{ij}^{-2\sigma}}.
  \label{eqn:probability}
\end{equation}
The interactions $J_{ij}$ between a pair of spins $(i,j)$ are
independent Gaussian random variables with mean zero and standard
deviation unity, i.e:
\begin{equation}
  \left[ J_{ij} \right]_{\text{av}}=0 \qquad \text{and} \qquad \left[ J_{ij}^2 \right]_{\text{av}} = J^2 = 1.
  \label{eqn:vb-interaction}
\end{equation}
The Cartesian components $h_i^{\mu}$ of the on-site external field are
independent random variables drawn from a Gaussian distribution of
zero mean with each component having variance $h_r^2$. The detailed
prescription to generate such a lattice with long-range diluted
interactions is given in
references~\cite{PhysRevLett.101.107203,sharma2011phase,XY:23}.

\begin{figure*}[ht]
  \gdef\sig{0.600}\setvaluesfixedhr{\sig}
  \centering
  \includegraphics[width=\textwidth]{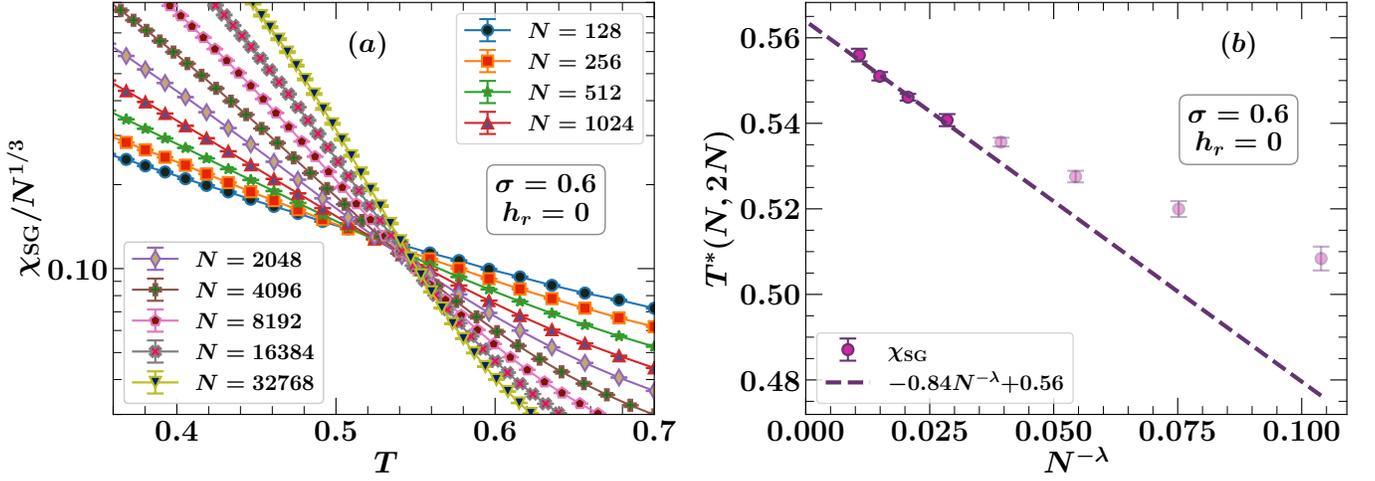}
  \caption{Finite size scaling analyses of data for $\sigma=\sig$
    obtained by varying the temperature in the absence of a magnetic
    field. (a) shows the plot of $\chi_{\text{SG}}/N^{1/3}$ as a
    function of the temperature $T$ for different system sizes. The
    plot shows that the curves for different system sizes
    intersect. The data for the intersection temperatures $T^*(N,2N)$
    between pairs of adjacent system sizes are plotted as a function
    of $N^{-\lambda}$ in (b). The value of the exponent $\lambda$ is
    fixed to be $\lam$ which is known exactly in the mean-field
    regime~\cite{sharma2011phase,PhysRevB.81.064415}. We fitted the
    $T^*(N,2N)$ data with
    Eq.~(\ref{eq:intersection_temperature_corrections}) using linear
    fitting and the resulting value of the transition temperature is
    $T_c = \chiTAT \pm \chiTATerror$ (see Table
    \ref{tab:results_fixed_hr} for details). The blurred points in (b)
    are excluded in from the linear fitting.}
  \label{fig:chi_vs_T_s0.600}
\end{figure*}

\begin{figure*}[t]
  \gdef\sig{0.600}\setvaluesfixedT{\sig}
  \centering
  \includegraphics[width=\textwidth]{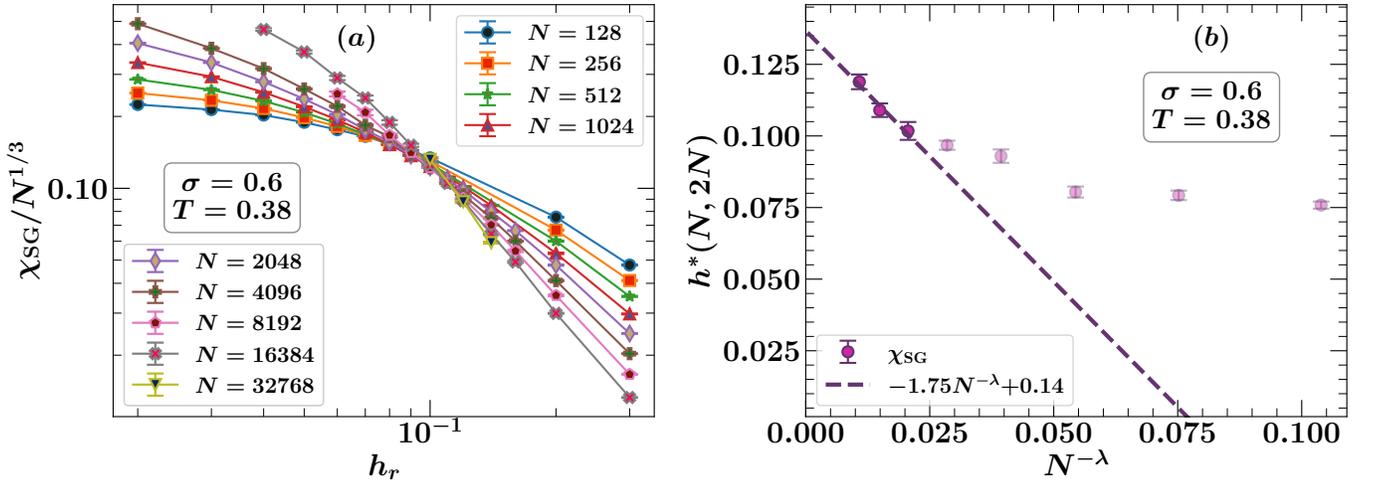}
  \caption{(a) Finite size scaling analyses of $\chi_{\text{SG}}$ data
    for $\sigma=\sig$ obtained by fixing the temperature to
    $T=\T\,(=\TbyTc\,T_c)$ and varying the field. The plot shows that
    the curves for different system sizes intersect. (b) shows the
    data for the intersection fields $h^*(N,2N)$ between pairs of
    adjacent system sizes, plotted as a function of
    $N^{-\lambda}$. Using $\lambda=\lam$ we fitted the $h^*(N,2N)$
    data linearly with Eq.~(\ref{eq:intersection_field_corrections})
    and the value of the transition field so obtained is
    $h_{\text{AT}}(T=\T)= \chihAT \pm \chihATerror$ (see Table
    \ref{tab:results_fixed_T} for details). The blurred points in (b)
    are excluded from the linear fitting.}
  \label{fig:chi_vs_hr_s0.600}
\end{figure*}

\begin{figure*}[ht]
  \gdef\sig{0.630}\setvaluesfixedhr{\sig}
  \centering
  \includegraphics[width=\textwidth]{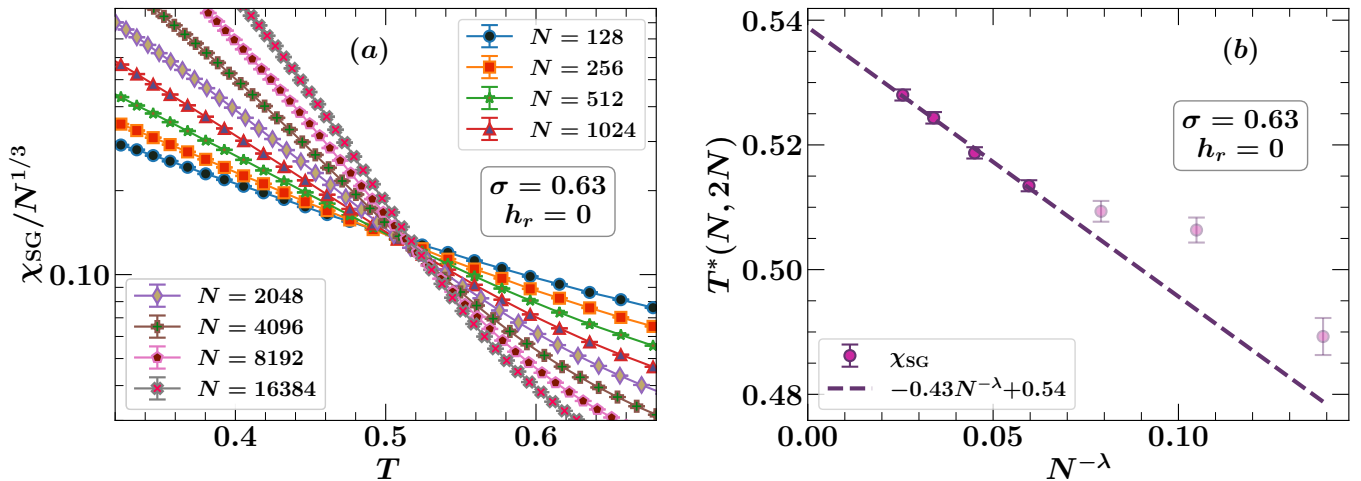}
  \caption{Finite size scaling analyses of data for $\sigma=\sig$
    obtained by varying the temperature in the absence of a magnetic
    field. (a) shows the plot of $\chi_{\text{SG}}/N^{1/3}$ as a
    function of the temperature $T$ for different system sizes. The
    data for the intersection temperatures $T^*(N,2N)$ between pairs
    of adjacent system sizes are plotted as a function of
    $N^{-\lambda}$ in (b). The value of the exponent $\lambda$ is
    fixed to be $\lam$. The line fit gives
    $T_c = \chiTAT \pm \chiTATerror$ (see Table
    \ref{tab:results_fixed_hr} for details). The blurred points in (b)
    are excluded from the linear fitting.}
  \label{fig:chi_vs_T_s0.630}
\end{figure*}

\begin{figure*}[ht]
  \gdef\sig{0.630}\setvaluesfixedT{\sig}
  \centering
  \includegraphics[width=\textwidth]{chi_vs_hr_s\sigint}
  \caption{(a) Finite size scaling analyses of $\chi_{\text{SG}}$ data
    for $\sigma=\sig$ obtained by fixing the temperature to
    $T=\T\,(=\TbyTc\,T_c)$ and varying the field. (b) shows the data
    for the intersection fields $h^*(N,2N)$ plotted as a function of
    $N^{-\lambda}$. Using $\lambda=\lam$ we fitted the $h^*(N,2N)$
    data linearly with Eq.~(\ref{eq:intersection_field_corrections})
    and the value of the transition field so obtained is
    $h_{\text{AT}}(T=\T)= \chihAT \pm \chihATerror$ (see Table
    \ref{tab:results_fixed_T} for details). The blurred points in (b)
    are excluded from the linear fitting.}
  \label{fig:chi_vs_hr_s0.630}
\end{figure*}

\begin{figure*}[ht]
  \gdef\sig{0.640}\setvaluesfixedhr{\sig}
  \centering
  \includegraphics[width=\textwidth]{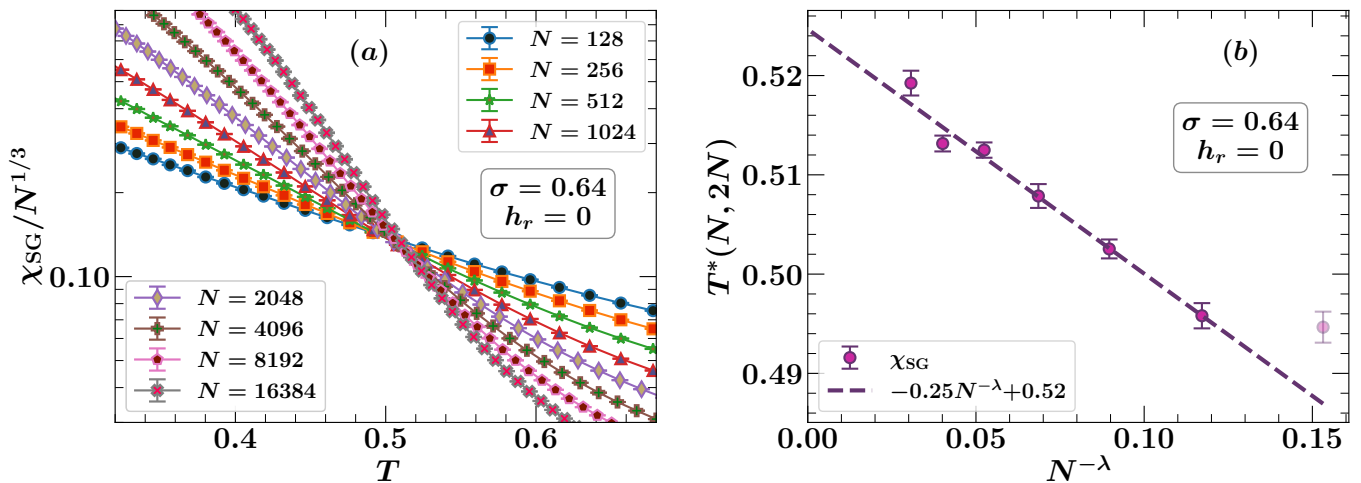}
  \caption{(a) Finite size scaling analyses of $\chi_{\text{SG}}$ data
    for $\sigma=\sig$ obtained by varying the temperature in the
    absence of a magnetic field. (b) shows the data for the
    intersection temperatures $T^*(N,2N)$ plotted as a function of
    $N^{-\lambda}$ with $\lambda=\lam$. The line fit gives
    $T_c = \chiTAT \pm \chiTATerror$ (see Table
    \ref{tab:results_fixed_hr} for details). The blurred point in (b)
    is excluded from the linear fitting.}
  \label{fig:chi_vs_T_s0.640}
\end{figure*}

\begin{figure*}[t]
  \gdef\sig{0.640}\setvaluesfixedT{\sig}
  \centering
  \includegraphics[width=\textwidth]{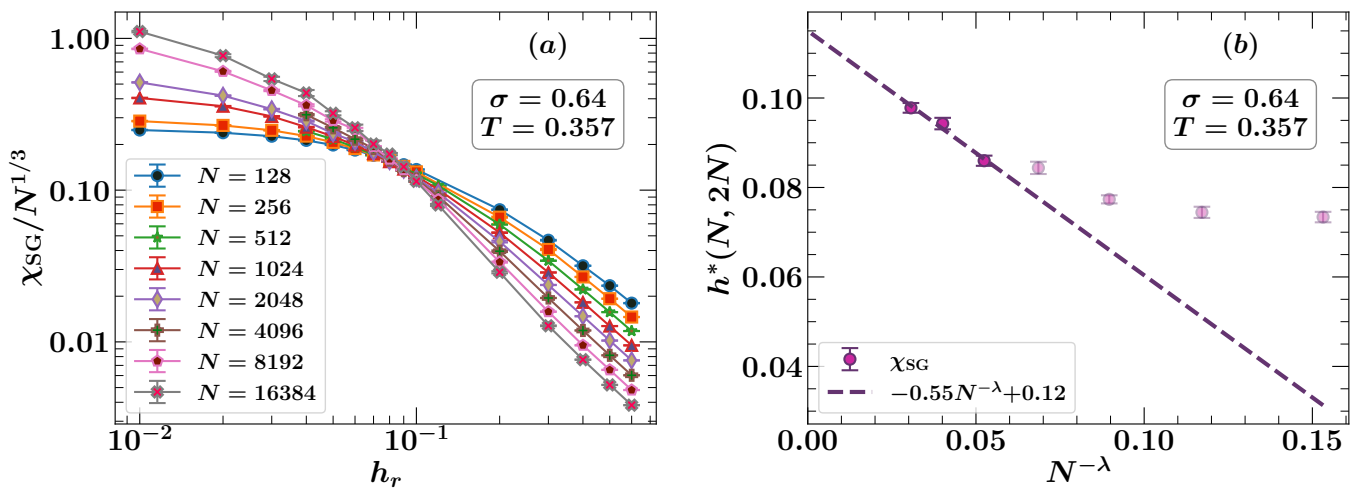}
  \caption{(a) Finite size scaling analyses of $\chi_{\text{SG}}$ data
    for $\sigma=\sig$ obtained by fixing the temperature to
    $T=\T\,(=\TbyTc\,T_c)$ and varying the field. (b) shows the data
    for the intersection fields $h^*(N,2N)$ between pairs of adjacent
    system sizes, fitted against $N^{-\lambda}$, using
    $\lambda=\lam$. The value of the transition field so obtained is
    $h_{\text{AT}}(T=\T)= \chihAT \pm \chihATerror$ (see Table
    \ref{tab:results_fixed_T} for details). The blurred points in (b)
    are excluded from the linear fitting.}
  \label{fig:chi_vs_hr_s0.640}
\end{figure*}

This model has already been extensively studied. Even though it
involves spins of $m$ $(=3)$ components, its AT transition is in the
universality class of the Ising ($m=1$) model
\cite{sharma2010almeida}. Despite the additional degrees of freedom of
the spins compared to those of the Ising model, the Heisenberg model
is easier to simulate than the Ising model as the vector spins provide
a means to go around barriers rather than over them as in the Ising
case, allowing larger systems to be simulated \cite{LeeYoung:07}.  In
the interval $1/2< \sigma < 2/3$, it corresponds to an
Edwards-Anderson short-range model in $d_{\text{eff}}$ dimensions
\cite{PhysRevB.67.134410}, where
\begin{equation}
  d_{\text{eff}}=\frac{2}{2\sigma-1}.
  \label{eq:deff}
\end{equation}
Thus if $\sigma=0.6$ (see Fig. \ref{fig:Heisenberg_AT}),
$d_{\text{eff}}=10$. We ourselves have extensively studied the XY
($m=2$) version of it \cite{XY:23}, when we concentrated mainly on
cases where $\sigma >2/3$. Since writing that paper we have discovered
that the Heisenberg case ($m=3$) runs faster, enabling us to study
larger systems. In this paper we have focussed on cases $\sigma< 2/3$
corresponding to $d > 6$ in an attempt to determine whether the AT
line vanishes as $d \to 6$. At the time of writing of our paper on the
XY spin glass model, we thought determining whether the AT line
vanished as $\sigma \to 2/3$ would be very challenging as the
corrections to scaling become larger and larger in this limit,
requiring the study of increasingly larger values of $N$ to achieve
the equivalent level of accuracy. Our work in this paper is indeed
affected by this difficulty which prevents us getting really close to
$\sigma = 2/3$ but it does suggest that the AT line might vanish at
$d=6$ (i.e. $\sigma = 2/3$) if the limit $N\to \infty$ could be
studied.

\begin{figure*}[ht]
  \gdef\sig{0.650}\setvaluesfixedhr{\sig}
  \centering
  \includegraphics[width=\textwidth]{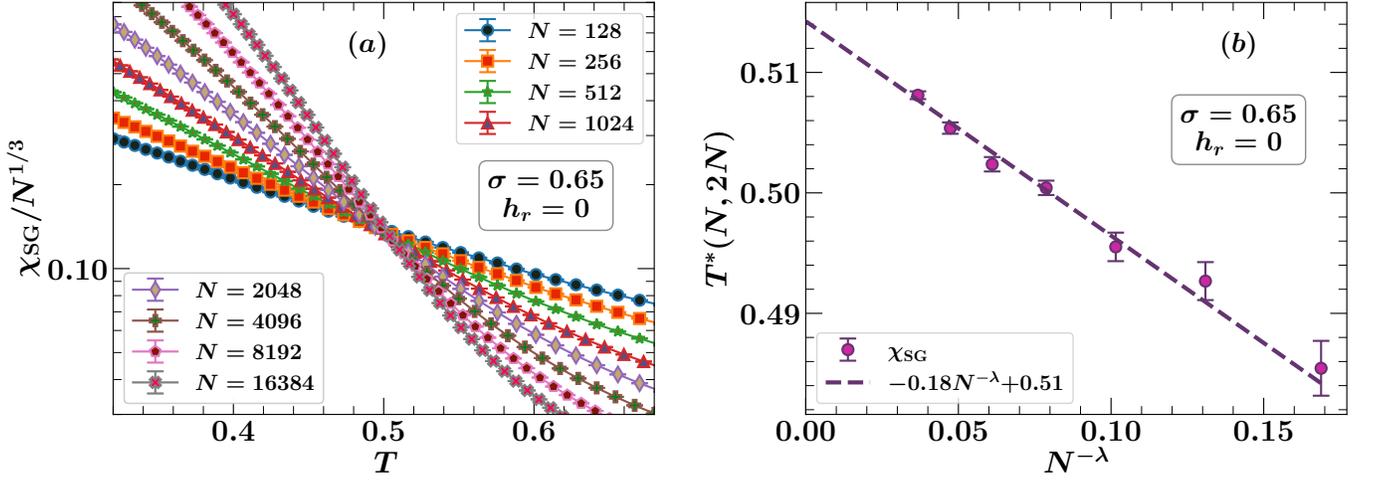}
  \caption{(a) Finite size scaling analyses of $\chi_{\text{SG}}$ data
    for $\sigma=\sig$ obtained by varying the temperature in the
    absence of a magnetic field. The data for the intersection
    temperatures $T^*(N,2N)$ are plotted as a function of
    $N^{-\lambda}$ (using $\lambda=\lam$) in (b). The value of the
    zero-field transition temperature obtained from linear fitting is
    $T_c = \chiTAT \pm \chiTATerror$ (see Table
    \ref{tab:results_fixed_hr} for details).}
  \label{fig:chi_vs_T_s0.650}
\end{figure*}

\begin{figure*}[t]
  \gdef\sig{0.650}\setvaluesfixedT{\sig}
  \centering
  \includegraphics[width=\textwidth]{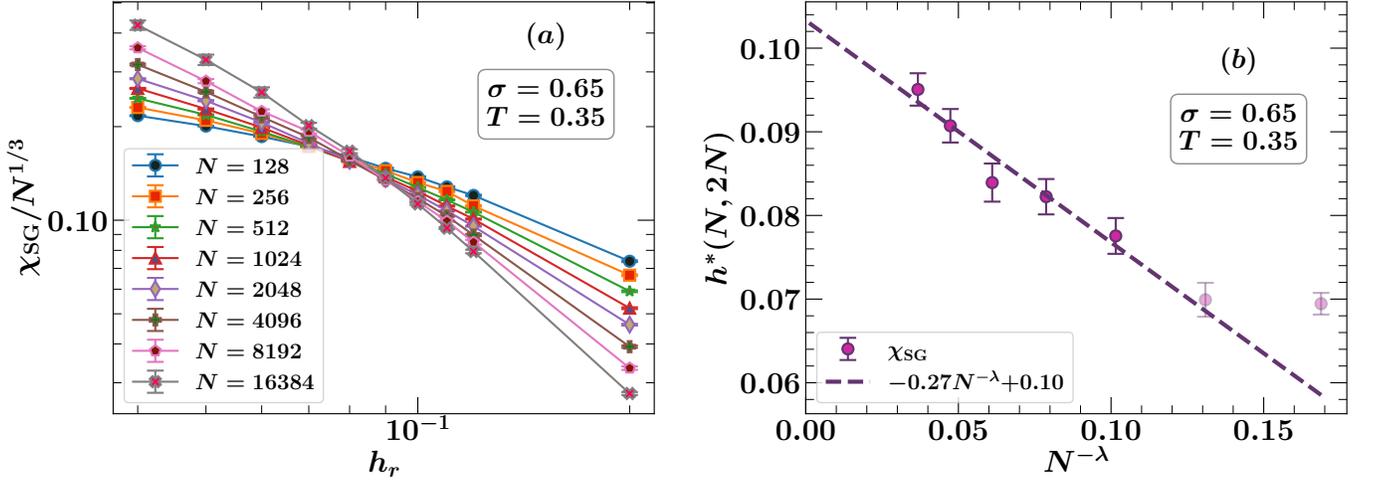}
  \caption{(a) Finite size scaling analyses of $\chi_{\text{SG}}$ data
    for $\sigma=\sig$ obtained by fixing the temperature to
    $T=\T\,(=\TbyTc\,T_c)$ and varying the field. (b) shows the data
    for the intersection fields $h^*(N,2N)$ between pairs of adjacent
    system sizes, plotted as a function of $N^{-\lambda}$. Using
    $\lambda=\lam$ we fitted the $h^*(N,2N)$ data linearly with
    Eq.~(\ref{eq:intersection_field_corrections}) and the value of the
    transition field so obtained is
    $h_{\text{AT}}(T=\T)= \chihAT \pm \chihATerror$ (see Table
    \ref{tab:results_fixed_T} for details). The blurred points in (b)
    are excluded from the linear fitting.}
  \label{fig:chi_vs_hr_s0.650}
\end{figure*}

\section{Correlation lengths and susceptibilities}
\label{sec:chi-xi}

As in the XY case we shall focus on the wave-vector-dependent spin
glass susceptibility \cite{sharma2011phase}
\begin{equation}
  \label{eq:wvd_sg_susceptibility}
  \chi_{\text{SG}}(k)=\frac{1}{N}\sum\limits_{i,j}\frac{1}{m}\sum\limits_{\mu,\nu}
  \left[\left(\chi_{ij}^{\mu\nu} \right)^2 \right]_{\text{av}}e^{ik(i-j)} ,
\end{equation}
where
\begin{equation}
  \label{eq:chi_ij^mu-nu}
  \chi_{ij}^{\mu\nu}=\left\langle S_i^{\mu} S_j^{\nu} \right\rangle -
  \left\langle S_i^{\mu} \right\rangle\left\langle S_j^{\nu} \right\rangle.
\end{equation}
From it the spin glass correlation length is then determined using the relation
\begin{equation}
  \label{eq:sg_correlation_length}
  \xi_{\text{SG}}=\frac{1}{2\sin(k_{\text{min}}/2)}
  \left(\frac{\chi_{\text{SG}}(0)}{\chi_{\text{SG}}\left( k_{\text{min}} \right)} -1 \right)^{1/(2\sigma-1)},
\end{equation}
and $k_{\text{min}}=2 \pi/N$. The spin glass susceptibility itself is
$\chi_{\text{SG}}=\chi_{\text{SG}}(0)$. The simulations and
equilibration checks were performed according to the procedures
outlined in Ref. \cite{LeeYoung:07,sharma2010almeida}, with the
details provided in the supplementary material (Section I).

At the AT transition, both $\chi_{\text{SG}}$ and $\xi_{\text{SG}}$
diverge to infinity. For $\sigma < 2/3$ the finite size scaling form
when approaching the AT line along a vertical trajectory (i.e. by
varying $h_r$) for a finite value of $N$ \cite{XY:23} is:
\begin{eqnarray}
  \frac{\chi_{\text{SG}}}{N^{1/3}}&=&\mathcal{C}\left[ N^{1/3}\left(h_r-h_{\text{AT}}(T)\right)\right] \nonumber \\
                                  &+& N^{-\omega}\mathcal{G}\left[ N^{1/3}\left(h_r-h_{\text{AT}}(T)\right)\right].
                                      \label{eq:scalingcorrform1}
\end{eqnarray}
The second term is a correction to scaling term. The exponent $\omega$
is given by \cite{sharma2010almeida,larson:2010}
\begin{equation}
  \omega=1/3-(2\sigma-1).
  \label{eq:omega?}
\end{equation}
Notice that as $\sigma \to 2/3$, $\omega\to 0$. This is why it is so
challenging to show that the AT line disappears as $d \to 6$.  The
finite size scaling form for $\xi_{\text{SG}}$ is \cite{XY:23}
\begin{eqnarray}
  \frac{\xi_{\text{SG}}}{N^{d_{\text{eff}}/6}} &= &\mathcal{X}\left[ N^{1/3}\left(h_r-h_{\text{AT}}(T)\right)\right]\nonumber\\
                                               &+& N^{-\omega}\mathcal{H}\left[N^{1/3}\left(h_r-h_{\text{AT}}(T)\right)\right].
                                                   \label{eq:xiscalingform}
\end{eqnarray}

In the absence of the correction to scaling term the plots of
$\chi_{\text{SG}}/N^{1/3}$ or $\xi_{\text{SG}}/N^{d_{\text{eff}}/6}$
for different system sizes would intersect at
$h_r=h_{\text{AT}}(T)$. The intersection formula for the successive
crossing points $h^*(N, 2N)$ should be linear in $1/N^{\lambda}$ when
$N\to \infty$ and be of the form
\begin{equation}
  h^*(N, 2N)=h_{\text{AT}}(T) +\frac{A}{N^{\lambda}},
  \label{eq:intersection_field_corrections}
\end{equation}
where 
\begin{equation}
  \lambda=1/3+\omega.
  \label{def:lambda}
\end{equation}

\begin{figure*}[ht]
  \gdef\sig{0.655}\setvaluesfixedhr{\sig}
  \centering
  \includegraphics[width=\textwidth]{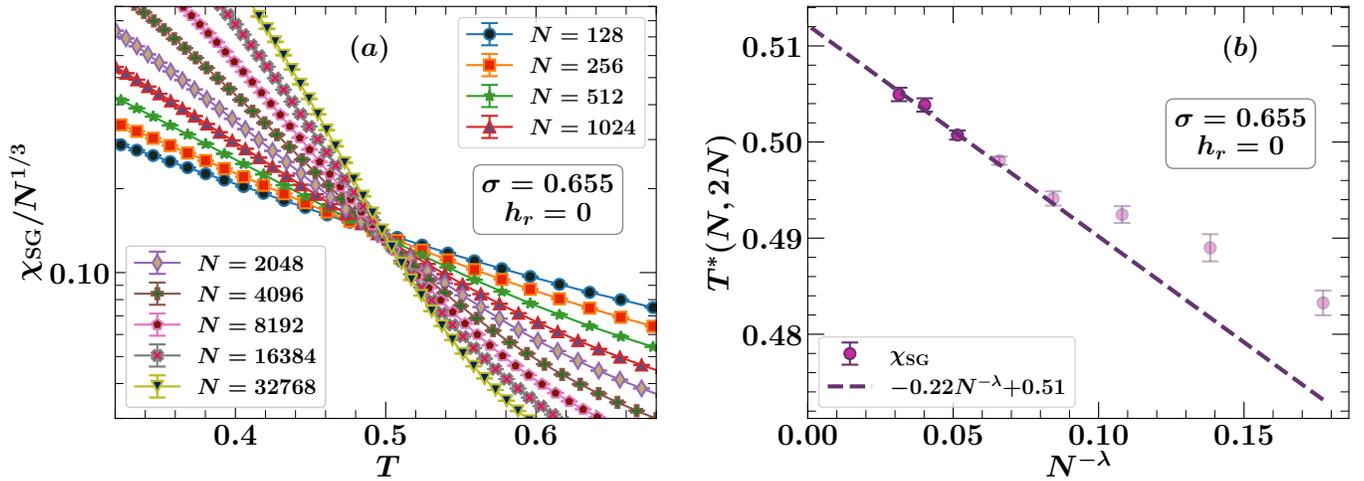}
  \caption{(a) Finite size scaling analyses of $\chi_{\text{SG}}$ data
    for $\sigma=\sig$ obtained by varying the temperature in the
    absence of a magnetic field. (b) shows the data for the
    intersection temperatures $T^*(N,2N)$ fitted against
    $N^{-\lambda}$. The value of the exponent $\lambda$ is fixed to be
    $\lam$. The line fit gives $T_c = \chiTAT \pm \chiTATerror$ (see
    Table \ref{tab:results_fixed_hr} for details). The blurred points
    in (b) are excluded from the linear fitting.}
  \label{fig:chi_vs_T_s0.655}
\end{figure*}

\begin{figure*}[t]
  \gdef\sig{0.655}\setvaluesfixedT{\sig}
  \centering
  \includegraphics[width=\textwidth]{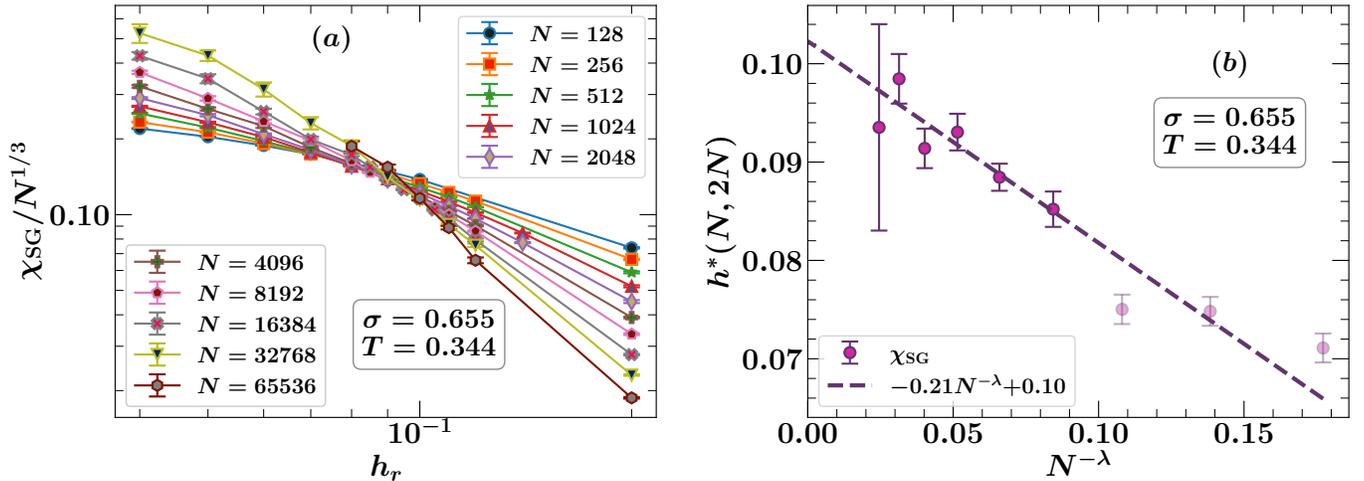}
  \caption{(a) Finite size scaling analyses of $\chi_{\text{SG}}$ data
    for $\sigma=\sig$ obtained by fixing the temperature to
    $T=\T\,(=\TbyTc\,T_c)$ and varying the field. In (b) the data for
    the intersection fields $h^*(N,2N)$ is plotted as a function of
    $N^{-\lambda}$ with $\lambda=\lam$. We fitted the $h^*(N,2N)$ data
    with a straight line and the value of the transition field
    obtained as a result of the extrapolation of the straight line is
    $h_{\text{AT}}(T=\T)= \chihAT \pm \chihATerror$ (see Table
    \ref{tab:results_fixed_T} for details). The blurred points in (b)
    are excluded from the linear fitting.}
  \label{fig:chi_vs_hr_s0.655}
\end{figure*}

We have not only studied $\chi_{\text{SG}}$ and $\xi_{\text{SG}}$ as a
function of $h_r$ at fixed $T$ but we have also studied them as a
function of $T$ for fixed $h_r= 0$.  We did the latter to determine
the zero-field transition temperature $T_c$.  The relevant finite size
scaling forms for this situation are
\begin{equation}
  \frac{\chi_{\text{SG}}}{N^{1/3}}=\mathcal{\tilde{C}}\left[ N^{1/3}(T-T_c)\right]  +
  N^{-\omega}\mathcal{\tilde{G}}\left[ N^{1/3}(T-T_c)\right],
\label{eq:scalingcorrformThr0}
\end{equation}
and 
\begin{equation}
  \frac{\xi_{\text{SG}}}{N^{d_{\text{eff}}/6}}=\mathcal{\tilde{X}}\left[ N^{1/3}(T-T_c)\right]  +
  N^{-\omega}\mathcal{\tilde{H}}\left[ N^{1/3}(T-T_c)\right].
\label{eq:xiscalingformThr0}
\end{equation}
The tilde sign is to indicate that the finite size scaling functions
such as $\mathcal{C}$ in a field and $\mathcal{\tilde{C}}$ in the
absence of a field may differ.

The intersection points as in, say, Fig. \ref{fig:chi_vs_T_s0.630}
\flc{b} would be expected to be of the form
\begin{equation}
T^*(N, 2N)=T_c + \frac{\tilde{A}}{N^{\lambda}}.
\label{eq:intersection_temperature_corrections}
\end{equation}

Note that the values of the exponents $\omega$ and $\lambda$ are the
same for the zero field transition and for the AT line (assuming that
they are both governed by a Gaussian fixed point).

In section \ref{fss-analyses} we give the results of our studies of
$\chi_{\text{SG}}$ for values of $\sigma$ at 0.600, 0.630, 0.640,
0.650 and 0.655. We also describe how the zero-field transition
temperature $T_c$ was determined for each of these values of
$\sigma$. The results obtained from $\xi_{\text{SG}}$ corresponding to
the same values of $\sigma$ are given in the supplementary material
(Section III).

Leuzzi et al. \cite{Leuzzi:09} pointed out that determination of
$\xi_{\text{SG}}$ was especially badly affected by finite size
effects. They found that if it were calculated not from the two $k$
values which we used, $0$ and $k_{\text{min}}$, but instead two
non-zero values, $k_1$ and $k_2$, a different value for
$\xi_{\text{SG}}$ was obtained. In a recent paper Aguilar-Janita et
al.  \cite{Martin-Mayor:23} pointed out that in zero field the two
methods gave the same result, and only differed in a field. They
attributed the difference to the fact that when studying the AT line
one requires the field to be large enough so that the Parisi overlap
function $P(q)$ vanishes for $q <0$. This requires
$\sqrt{q} h_r \sqrt{N} >k_BT$. When this criterion is not satisfied
there are additional crossovers and length scales beyond those
discussed in this section. As we find that the field at the AT line is
approaching zero as $\sigma \to 2/3$ the criterion requires the study
of even larger system sizes as $\sigma \to 2/3$, which makes our
results from $\xi_{\text{SG}}$ especially unreliable and it is for
this reason we have relegated them to supplementary material (Section
III). Finite size effects affect some quantities more than others and
hopefully our estimates of $h_{\text{AT}}$ from $\chi_{\text{SG}}$ are
less affected. This might be because the values of
$\chi_{\text{SG}}/N^{1/3}$ in zero field and on the AT line are rather
similar, whereas the values of $\xi_{\text{SG}}/N^{d_{\text{eff}}/6}$
on the AT line are larger than their values in zero field by a factor
of order $10$.

\section{Finite size scaling analyses}
\label{fss-analyses}

\begin{table*}[t]
  \centering    
  \caption{Results of the simulations done by varying the temperture
    $T$ in the absence of magnetic field $h_r$. The
    $\chi_{\text{SG}}/N^{1/3}$ when plotted as a function of
    temperature $T$, the data for different system sizes $N$ intersect
    around the transition temperature $T_c(\chi_{\text{SG}})$. The
    intersection temperatures $T^*(N,2N)$ between the curves for two
    adjacent system sizes are then plotted as a function of
    $N^{-\lambda}$, with $\lambda=5/3 - 2\sigma$ in the mean field
    regime~\cite{sharma2011phase,PhysRevB.81.064415}. We then fit this
    data for the $N_{\text{pairs}}$ largest pairs of system sizes with
    Eq. (\ref{eq:intersection_temperature_corrections}) to find the
    transition temperature $T_c(\chi_{\text{SG}})$. For linear
    fitting, we attempt to fit the intersection data with a straight
    line for various values of $N_{\text{pairs}}$, and we choose the
    best fit where $\chi^2/N_{\text{dof}}$ is closest to one.}
  \label{tab:results_fixed_hr}
  \begin{ruledtabular}
    \begin{tabular}{cccccc}
      $\sigma$  & $h_r$  & $\lambda$  & $N_{\text{pairs}}(\chi_{\text{SG}})$  & $T_{c}(\chi_{\text{SG}})$  &  $\chi^2/N_{\text{dof}}$    \\[0.15cm]
      \hline
      \gdef\sig{0.600}\setvaluesfixedhr{\sig}\\[-0.45cm]
      \sig  &  \hr  &  \lam  &  \chinpairs  & $\chiTAT \pm \chiTATerror$  &  \chichisq   \\
      \gdef\sig{0.630}\setvaluesfixedhr{\sig}\\[-0.45cm]
      \sig  &  \hr  &  \lam  &  \chinpairs  & $\chiTAT \pm \chiTATerror$  &  \chichisq   \\
      \gdef\sig{0.640}\setvaluesfixedhr{\sig}\\[-0.45cm]
      \sig  &  \hr  &  \lam  &  \chinpairs  & $\chiTAT \pm \chiTATerror$  &  \chichisq   \\
      \gdef\sig{0.650}\setvaluesfixedhr{\sig}\\[-0.45cm]
      \sig  &  \hr  &  \lam  &  \chinpairs  & $\chiTAT \pm \chiTATerror$  &  \chichisq   \\
      \gdef\sig{0.655}\setvaluesfixedhr{\sig}\\[-0.45cm]
      \sig  &  \hr  &  \lam  &  \chinpairs  & $\chiTAT \pm \chiTATerror$  &  \chichisq   \\
    \end{tabular}
  \end{ruledtabular}   
\end{table*}

\begin{table*}[t]
   \centering    
   \caption{Results of the simulations done by varying the magnetic
     field $h_r$ at a fixed temperature $T$.  $T_c$ is the zero-field
     spin glass transition temperature obtained from the last column
     of the Table \ref{tab:results_fixed_hr}. Similiar to the fixed
     $h_r$ case described in Table~\ref{tab:results_fixed_hr}, we plot
     the finite-size-scaled $\chi_{\text{SG}}$ data as a function of
     the field $h_r$ and find the intersection fields $h^*(N,2N)$. We
     then fit this data for the $N_{\text{pairs}}$ largest pairs of
     system sizes with Eq.~(\ref{eq:intersection_field_corrections})
     to find the AT transition field $h_{\text{AT}}$ corresponding to
     the temperature $T$. For linear fitting, we attempt to fit the
     intersection data with a straight line for various values of
     $N_{\text{pairs}}$, and we choose the best fit where
     $\chi^2/N_{\text{dof}}$ is closest to one.}
   \label{tab:results_fixed_T}
   \begin{ruledtabular}
     \begin{tabular}{cccccccc}
       $\sigma$  & $T$   & $T_c$  &  $T/T_c$   & $\lambda$   & $N_{\text{pairs}}(\chi_{\text{SG}})$   & $h_{\text{AT}}(\chi_{\text{SG}})$  &  $\chi^2/N_{\text{dof}}$   \\[0.15cm]
       \hline
       \gdef\sig{0.600}\setvaluesfixedT{\sig}\\[-0.45cm]
       \sig  &  \T  &  \Tc  &  \TbyTc  &  \lam  &  \chinpairs  & $\chihAT \pm \chihATerror$  &  \chichisq   \\
       \gdef\sig{0.630} \setvaluesfixedT{\sig}\\[-0.45cm]
       \sig  &  \T  &  \Tc  &  \TbyTc  &  \lam  &  \chinpairs  & $\chihAT \pm \chihATerror$  &  \chichisq   \\
       \gdef\sig{0.640} \setvaluesfixedT{\sig}\\[-0.45cm]
       \sig  &  \T  &  \Tc  &  \TbyTc  &  \lam  &  \chinpairs  & $\chihAT \pm \chihATerror$  &  \chichisq   \\
       \gdef\sig{0.650} \setvaluesfixedT{\sig}\\[-0.45cm]
       \sig  &  \T  &  \Tc  &  \TbyTc  &  \lam  &  \chinpairs  & $\chihAT \pm \chihATerror$  &  \chichisq   \\
       \gdef\sig{0.655} \setvaluesfixedT{\sig}\\[-0.45cm]
       \sig  &  \T  &  \Tc  &  \TbyTc  &  \lam  &  \chinpairs  & $\chihAT \pm \chihATerror$  &  \chichisq   \\
     \end{tabular}
   \end{ruledtabular}   
\end{table*}

Here we give further details of the results of our simulations for
$\sigma$ values 0.600, 0.630, 0.640, 0.650 and
0.655. Figs. \ref{fig:chi_vs_T_s0.600}, \ref{fig:chi_vs_T_s0.630},
\ref{fig:chi_vs_T_s0.640}, \ref{fig:chi_vs_T_s0.650}, and
\ref{fig:chi_vs_T_s0.655} show the finite size scaling analyses of
data for these different values of $\sigma$ obtained by varying the
temperature in the absence of a magnetic field. In all these figures,
Figs. (a) show the plot of $\chi_{\text{SG}}/N^{1/3}$ as a function of
the temperature $T$ for different system sizes. In all these sets of
plots we can clearly notice that the curves for different system sizes
intersect around the transition temperature, which is in accordance
with Eq.~(\ref{eq:scalingcorrformThr0}). For each pair of adjacent
system sizes, we find the intersection temperature $T^*(N,2N)$ from
$\chi_{\text{SG}}$ data which is the $x-$ coordinate corresponding to
the point of intersection between these curves. The data for the
intersection temperatures obtained from all the pairs of adjacent
system sizes are plotted as a function of $N^{-\lambda}$ in
Figs. (b). The value of the exponent $\lambda$ is known in the
mean-field regime and is given by
Eq. (\ref{def:lambda})~\cite{sharma2011phase,PhysRevB.81.064415}. Using
this value of $\lambda$ we fit the $T^*(N,2N)$ data linearly with
Eq. (\ref{eq:intersection_temperature_corrections}) for the
$N_{\text{pairs}}$ largest pairs of system sizes. Throughout the
paper, the fitting procedure is carried out using the
\href{https://docs.scipy.org/doc/scipy/reference/generated/scipy.optimize.curve_fit.html}{\textit{curve\_fit}}
function from the Python library SciPy~\cite{2020SciPy-NMeth}. This
function employs the Levenberg-Marquardt (LM)
algorithm~\cite{press2009numerical,LM_algorithm} to fit the data,
providing both the fitting parameters and their corresponding
errors. In the thermodynamic limit $N^{-\lambda} \to 0$ as
$N \to \infty$. Hence, the $y$--intercept corresponding to the
straight line fit gives us the value of the zero-field transition
temperature $T_c$. The values of $T_c$ obtained for different values
of $\sigma$ are shown in Table \ref{tab:results_fixed_hr}, and the
parameters of the simulations are shown in Table S1 of the
supplementary material.

The AT line can be approached not only by reducing the temperature $T$
but also by reducing the field at fixed $T$. This was the procedure
used in Ref.~\cite{XY:23}. These are the vertical trajectories in
Fig. \ref{fig:Heisenberg_AT} along which we can cross the AT
line. Throughout this paper, we chose the value of temperature $T$
such that $T/T_c \approx 0.67$. We show our finite size scaling
analyses plots corresponding to this procedure in
Figs. \ref{fig:chi_vs_hr_s0.600}, \ref{fig:chi_vs_hr_s0.630},
\ref{fig:chi_vs_hr_s0.640}, \ref{fig:chi_vs_hr_s0.650}, and
\ref{fig:chi_vs_hr_s0.655}. Similar to the zero-magnetic field case,
we present our $\chi_{\text{SG}}$ data in Figs. (a) as a function of
magnetic field $h_r$. According to Eq.  (\ref{eq:scalingcorrform1}),
the data for $\chi_{\text{SG}}/N^{1/3}$ when plotted for different
system sizes should intersect at the AT transition field
$h_{\text{AT}}(T)$. Figs. (b) show the data for the intersection
fields $h^*(N,2N)$ obtained by considering the curves for adjacent
system sizes. We fit the $h^*(N,2N)$ data with
Eq. (\ref{eq:intersection_field_corrections}) through a straight line
for the $N_{\text{pairs}}$ largest pairs of system sizes using the
same value of $\lambda$ as in the previous scenario, which is given by
Eq. (\ref{def:lambda}). The point at which this straight line cuts the
$y-$ axis gives us the value of the transition field $h_{\text{AT}}$
corresponding to the temperature $T$. The values of $h_{\text{AT}}$
obtained for different $\sigma$ are shown in Table
\ref{tab:results_fixed_T}, and the parameters of the simulation are
shown in Table S2 of the supplementary material.

We have also performed finite size scaling analyses on
$\xi_{\text{SG}}$ data for all the 5 values of $\sigma$ both in the
zero field case and by varying the field at a fixed
temperature. Similar to $\chi_{\text{SG}}$ we plotted
$\xi_{\text{SG}}/N^{d_{\text{eff}}/6}$ for different system sizes as a
function of temperature $T$ for $h_r=0$, with
$d_{\text{eff}}=2/(2\sigma-1)$, and obtained the values of
intersection temperatures $T^*(N,2N)$. As for the vertical trajectory,
we fixed $T\approx 0.67\,T_c$ and, plotted
$\xi_{\text{SG}}/N^{d_{\text{eff}}/6}$ as a function of field $h_r$,
and computed the values of intersection fields $h^*(N,2N)$. We then
analysed the intersection temperatures or fields data as a function of
$N^{-\lambda}$ with $\lambda$ being the same for both $T^*$ and $h^*$
data sets obtained from both $\chi_{\text{SG}}$ and
$\xi_{\text{SG}}$. Upon doing linear fitting we obtained the values of
$T_c$ and $h_{\text{AT}}(T)$ from the $T^*$ and $h^*$ data sets
respectively. In the following sections we present the results of
finite size scaling analyses on $\chi_{\text{SG}}$ data for different
values of $\sigma$.

\subsection{$\sigma=0.600$}
\label{sec:sigma_0.600}

\setvaluesfixedT{0.600} At $\sigma=0.600$, for which
$d_{\text{eff}}= 10$ the results should be quite close to those of the
SK model (but see Fig.~\ref{fig:Heisenberg_AT}): it is in the same
mean-field regime and the exponent $\zeta =3$. The zero-field
transitions for this case has been studied by one of the authors of
this paper in Ref.~\cite{sharma2011phase} for Heisenberg spins. As a
sanity check, we attempted to replicate this analysis, and the
results, displayed in Fig. \ref{fig:chi_vs_T_s0.600}, are in complete
agreement with those presented in Ref.~\cite{sharma2011phase}. The
value of the zero-field spin glass transition temperature found from
these simulations is $T_c=\Tc$. The phase transitions in the presence
of an external magnetic field has also been studied in
Ref.~\cite{sharma2011almeida} for $h_r=0.1$. It has been reported that
the system undergoes a phase transition at
$T_{\text{AT}}(h_r=0.1)=0.406$.

\setvaluesfixedT{0.600} For $\sigma = 0.600$ we fixed the temperature
at $T=\T\, (=\TbyTc \, T_c)$. We have constructed the crossing plots
for $\chi_{\text{SG}}$ as a function of $h_r$ in
Fig. \ref{fig:chi_vs_hr_s0.600}\flc{a}. Analysis of the crossing
points $h^*(N, 2N)$ in Fig. \ref{fig:chi_vs_hr_s0.600}\flc{b} shows
that the behavior is again consistent with the existence of an AT line
at least at $\sigma = 0.600$. The value of the exponent $\lambda$ is
known in the mean-field regime and is given by
$\lambda=5/3-2\sigma=\lam$. The $h^*(N,2N)$ data for the largest
$\chinpairs$ pairs of system sizes are fitted against $N^{-\lambda}$
to give $h_{\text{AT}}(T=\T) = \chihAT \pm \chihATerror$ (see Table
\ref{tab:results_fixed_T}).

\subsection{$\sigma=0.630$}
\label{sec:sigma_0.630}

\setvaluesfixedhr{0.630} For $\sigma=0.630$
$d_{\text{eff}}\approx 7.692$. Our results for $h_r=0$ are given in
Fig. \ref{fig:chi_vs_T_s0.630}. According to
Eq. (\ref{eq:scalingcorrformThr0}), the data for
$\chi_{\text{SG}}/N^{1/3}$ when plotted for different system sizes
should intersect at the transition temperature
$T_c$. Fig. \ref{fig:chi_vs_T_s0.630}\flc{a} shows the data for
different system sizes.  We find the temperature $T^*(N,2N)$ at which
the curves corresponding to the system sizes $N$ and $2N$
intersect. We then fit this data with
Eq. (\ref{eq:intersection_temperature_corrections}) to find the
transition temperature.  The exponent $\lambda \equiv 5/3-2\sigma$ is
known to equal $\lam$ in this case.  The result is displayed in
Fig.~\ref{fig:chi_vs_T_s0.630}\flc{b}, where the $T^*(N,2N)$ data
obtained from intersections of $\chi_{\text{SG}}$ are fitted against
$N^{-\lambda}$ with a straight line for the largest $\chinpairs$ pairs
of system sizes to give $T_{c}= \chiTAT \pm \chiTATerror$ (see Table
\ref{tab:results_fixed_hr}).

\setvaluesfixedT{0.630} We have also studied $\chi_{\text{SG}}$ at
fixed $T$, but varying $h_r$ and the finite size scaling plots for
these are given in Fig. \ref{fig:chi_vs_hr_s0.630}\flc{a}. There
appears to be good intersections in the curves, supporting therefore
the possible existence of an AT transition at the temperature studied
$T = \T$. A plot of $h^*(N, 2N)$ versus $1/N^{\lambda}$ is in
Fig. \ref{fig:chi_vs_hr_s0.630}\flc{b}, using the same value of
$\lambda=\lam$. Considering the data for the $\chinpairs$ largest
pairs of system sizes, we did a linear fitting over the $h^*(N,2N)$
data obtained from $\chi_{\text{SG}}$ intersections, which gives
$h_{\text{AT}}(T=\T) = \chihAT \pm \chihATerror$ (see Table
\ref{tab:results_fixed_T}).

\subsection{$\sigma=0.640$}
\label{sec:sigma_0.640}

\setvaluesfixedhr{0.640} For $\sigma=0.640$,
$d_{\text{eff}}\approx 7.143$. Our results for $h_r=0$ are given in
Fig.~\ref{fig:chi_vs_T_s0.640}. Fig. \ref{fig:chi_vs_T_s0.640}\flc{a}
shows the $\chi_{\text{SG}}$ data for different system sizes, and the
corresponding intersection temperatures data are displayed in
Fig. \ref{fig:chi_vs_T_s0.640}\flc{b}. The value of the exponent
$\lambda$ for this case is $\lam$. The linear fit over the $T^*(N,2N)$
data obtained from intersections of $\chi_{\text{SG}}$, considering
the $\chinpairs$ pairs of largest system sizes, gives
$T_{c}= \chiTAT \pm \chiTATerror$ (see Table
\ref{tab:results_fixed_hr}).

\setvaluesfixedT{0.640} As for the alternate protocol where we fix the
temperature and vary the field, the finite size scaling plots are
given in Fig. \ref{fig:chi_vs_hr_s0.640}\flc{a}. The temperature is
fixed at $T = \T \, (=\TbyTc \, T_c)$. A plot of $h^*(N, 2N)$ versus
$1/N^{\lambda}$ is in Fig. \ref{fig:chi_vs_hr_s0.640}\flc{b}, using
the same value of $\lambda=\lam$. Omitting the smallest system size,
we did a linear fitting over the $h^*(N,2N)$ data obtained from
$\chi_{\text{SG}}$ intersections, which gives
$h_{\text{AT}}(T=\T) = \chihAT \pm \chihATerror$ (see Table
\ref{tab:results_fixed_T}).

\subsection{$\sigma=0.650$}
\label{sec:sigma_0.650}

\setvaluesfixedhr{0.650} For $\sigma=0.650$
$d_{\text{eff}}\approx 6.667$. Our results for $h_r=0$ are given in
Fig.~\ref{fig:chi_vs_T_s0.650}. The $\chi_{\text{SG}}$ data for
different system sizes are shown in
Fig. \ref{fig:chi_vs_T_s0.650}\flc{a}.  We find the intersection
temperatures $T^*(N,2N)$ and fit this data with
Eq. (\ref{eq:intersection_temperature_corrections}) to find the
transition temperature.  The exponent $\lambda \equiv 5/3-2\sigma$ is
known to equal $\lam$ in this case
~\cite{sharma2011phase,PhysRevB.81.064415}.  The result is displayed
in Fig. \ref{fig:chi_vs_T_s0.650}\flc{b}, where the $T^*(N,2N)$ data
obtained from intersections of $\chi_{\text{SG}}$ are fitted against
$N^{-\lambda}$ with a straight line for the largest $\chinpairs$ pairs
of system sizes to give $T_{c}= \chiTAT \pm \chiTATerror$ (see Table
\ref{tab:results_fixed_hr}).

\setvaluesfixedT{0.650} We have also studied $\chi_{\text{SG}}$ at
fixed temperature $T=\T \, (=\TbyTc\, T_c)$, but varying $h_r$ and the
finite size scaling plots for these are given in
Fig. \ref{fig:chi_vs_hr_s0.655}\flc{a}. A plot of $h^*(N, 2N)$ versus
$1/N^{\lambda}$ is in Fig. \ref{fig:chi_vs_hr_s0.650}\flc{b}, using
the same value of $\lambda=\lam$. Omitting the smallest system size,
the linear fit from $\chi_{\text{SG}}$ intersections give
$h_{\text{AT}}(T=\T) = \chihAT \pm \chihATerror$ (see Table
\ref{tab:results_fixed_T}).

\begin{figure}
  \centering
  \includegraphics[width=0.48\textwidth]{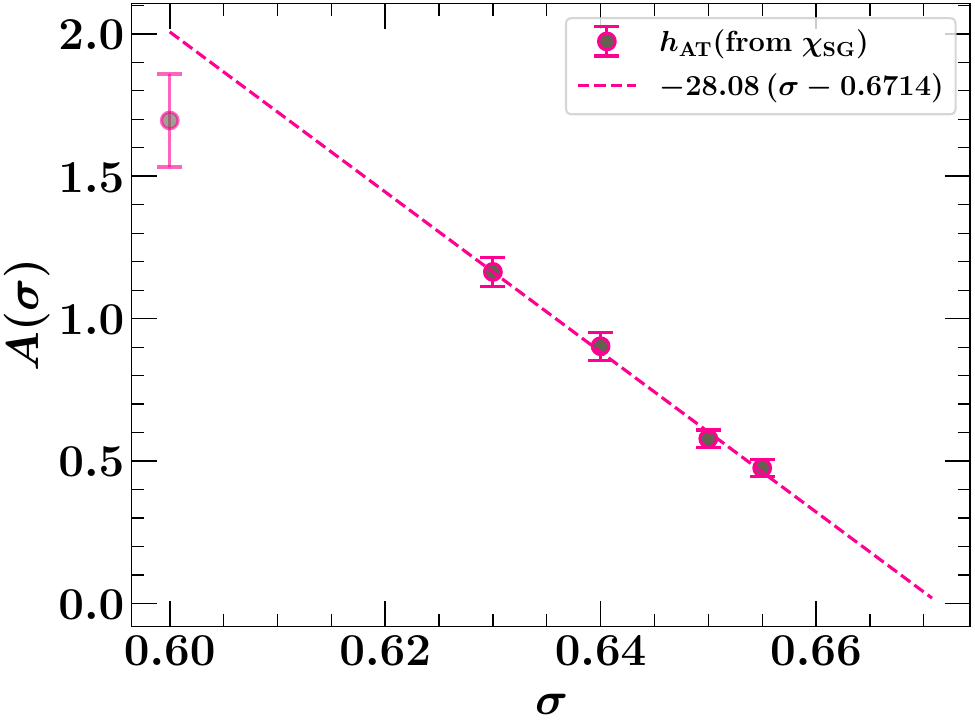}
  \caption{Plot of $ A(\sigma) $ versus $ \sigma $. The quantity
      $ A(\sigma) $ is computed using Eq.~(\ref{eq:approx_ATMFT}),
      with the exponent $ \zeta $ given by Eq.~(\ref{eq:xval}) in the
      mean-field regime. In our simulations, we fix the temperature at
      \( T \approx 0.67 \, T_c \) and determine the transition field
      \( h_{\text{AT}} \) from the \( \chi_{\text{SG}} \) dataset
      (see, for example, Fig.~\ref{fig:chi_vs_hr_s0.630}). The
      $ A(\sigma) $ data, excluding the point at $ \sigma = 0.600 $,
      are fitted with a straight line, yielding a reduced chi-squared
      value of $ \chi^2 / N_{\text{dof}} = 0.498 $. The slope and
      $ y $-intercept of this fit are $ m = -28.082 \pm 2.175 $ and
      $ c = 18.856 \pm 1.410 $, respectively. The correlation
      coefficient between $m$ and $c$ is $\rho_{mc} \approx -1.0$,
      indicating a strong negative correlation, which means an
      increase in the slope $m$ is accompanied by a nearly
      proportional decrease in the intercept $c$. The line intersects
      the $\sigma$-axis at $\sigma_l = -c/m = 0.671 \pm 0.002$. This
      corresponds to $d_l = 5.833 \pm 0.065$. In these linear fits,
      the data point at $ \sigma = 0.600 $ is excluded (and blurred)
      because it lies outside the linear region, which is valid only
      for $ \sigma $ values near $ 2/3 $. The error bars shown are
      statistical, but finite-size effects introduce an unknown
      systematic error in all data points. Details of the error
      calculations for $ A(\sigma) $, $ \sigma_l $, and $ d_l $ are
      provided in the supplementary material.  }
  \label{fig:A_vs_sigma}
\end{figure}

\subsection{$\sigma=0.655$}
\label{sec:sigma_0.655}

\setvaluesfixedhr{0.655} For $\sigma=0.655$
$d_{\text{eff}}\approx 6.452$. Our results for $h_r=0$ are given in
Fig.~\ref{fig:chi_vs_T_s0.655}. The $\chi_{\text{SG}}/N^{1/3}$ data
are plotted as a function of tempereture $T$ in
Fig. \ref{fig:chi_vs_T_s0.655}\flc{a} for different system sizes.  We
find the temperature $T^*(N,2N)$ at which the curves corresponding to
the system sizes $N$ and $2N$ intersect. We then fit this data with
Eq. (\ref{eq:intersection_temperature_corrections}) to find the
transition temperature.  The exponent $\lambda \equiv 5/3-2\sigma$ is
known to equal $\lam$ in this case
~\cite{sharma2011phase,PhysRevB.81.064415}.  The result is displayed
in Fig. \ref{fig:chi_vs_T_s0.655}\flc{b}, where the $T^*(N,2N)$ data
obtained from intersections of $\chi_{\text{SG}}$ are fitted against
$N^{-\lambda}$ with a straight line for the largest $\chinpairs$ pairs
of system sizes to give $T_{c}= \chiTAT \pm \chiTATerror$ (see Table
\ref{tab:results_fixed_hr}).

\setvaluesfixedT{0.655} We have also studied $\chi_{\text{SG}}$ at
fixed $T=\T \,(=\TbyTc \, T_c)$, but varying $h_r$ and the finite size
scaling plots for these are given in
Fig. \ref{fig:chi_vs_hr_s0.655}\flc{a}. A plot of $h^*(N, 2N)$ versus
$1/N^{\lambda}$ is in Fig. \ref{fig:chi_vs_hr_s0.655}\flc{b}, using
the same value of $\lambda=\lam$. Omitting the smallest system size,
we did a linear fitting over the $h^*(N,2N)$ data obtained from
$\chi_{\text{SG}}$ intersections, which gives
$h_{\text{AT}}(T=\T) = \chihAT \pm \chihATerror$ (see Table
\ref{tab:results_fixed_T}).


\section{Disappearance of the AT line as $\sigma \to 2/3$}
\label{sec:A_sigma_analysis}

In this section we present our analysis of $A(\sigma)$ for different
values of $\sigma<2/3$, and with the help of this data, we show that
the AT line approaches the horizontal axis as we go below six
dimensions, or equivalently for $\sigma>2/3$. As shown in Table
\ref{tab:results_fixed_T}, for each $\sigma$, we get two estimates of
the value of AT transition field $h_{\text{AT}}$ for a fixed
temperature $T$; one from $\chi_{\text{SG}}$ and one from
$\xi_{\text{SG}}$.  From it, one can extract (using
Eq. (\ref{eq:approx_ATMFT})) $A(\sigma)$ at the values of $\sigma$
which we have studied. We take the exponent to be:
\begin{equation}
  \zeta =
  \begin{cases}
    3, &\;\; \sigma < 5/8, \\[0.1cm]
    \dfrac{2(1-\sigma)}{2\sigma-1}, &\;\; 5/8 < \sigma < 2/3,
  \end{cases}
  \label{eq:xval}
\end{equation}
where we have set $d=d_{\text{eff}}$ in $\zeta=d/2-1$ for
$\sigma > 5/8$ \citep{Green:83,FisherSomp:85}. We have plotted the
results in Fig. \ref{fig:A_vs_sigma}, where $A(\sigma)$ clearly
decreases with increasing $\sigma$. In this linear plot, it appears to
approach zero at $\sigma_l = 0.671 \pm 0.002$, corresponding to
an effective spatial dimension of $d_l = 5.833 \pm 0.065$. This
value of $\sigma$ is close to $2/3$, which aligns with the expectation
that the AT line vanishes at exactly six dimensions.

\section{Summary and conclusions}
\label{sec:summary}

Due to the challenges associated with performing simulations above six
dimensions, we have opted to perform simulations using a
one-dimensional proxy model instead.  For the one dimensional
Heisenberg spin glasses with power-law diluted interactions, we
studied five values of $\sigma<2/3$: $0.600$, 0.630, 0.640, 0.650, and
0.655.

To determine the zero-field spin glass transition temperature $T_c$,
we performed simulations by varying the temperature $T$ in the absence
of a magnetic field. After estimating $T_c$, we fixed the temperature
at $T \approx 0.67 \, T_c$ and generated data by varying the magnetic
field for different values of $\sigma$. The largest system sizes we
studied include $N=32768$ for $\sigma=0.600 \text{ and } 0.630$,
$N=16384$ for $\sigma=0.640 \text{ and } 0.650$, and $N=65536$ for
$\sigma=0.655$. Using standard finite-size scaling analysis, we
identified the AT transition field $h_{\text{AT}}$, which allowed us
to compute the values of $A(\sigma)$.

Our results, as shown in Fig. \ref{fig:A_vs_sigma}, indicate that
\(A(\sigma)\) decreases as \(\sigma\) approaches \(2/3\),
corresponding to the effective spatial dimension of
\(d=6\). Specifically, we find that \(A(\sigma)\) becomes zero at
\(\sigma_l = 0.671 \pm 0.002\), which corresponds to
\(d_l = 5.833 \pm 0.065\). This implies that as the spatial
dimension approaches six, the AT line vanishes.


Numerical studies like this can only provide evidence for what the
truth might be: they do not as yet prove it beyond reasonable doubt.
The controversy will probably only be ended by a rigorous
determination of the lower critical dimension of the AT transition.

\section*{Acknowledgments}
We are grateful to the High Performance Computing (HPC) facility at
IISER Bhopal, where large-scale calculations in this project were
run. B.V
is grateful to the Council of Scientific and Industrial Research
(CSIR), India, for his PhD fellowship. A.S acknowledges financial
support from SERB via the grant (File Number: CRG/2019/003447), and
from DST via the DST-INSPIRE Faculty Award
[DST/INSPIRE/04/2014/002461].

\twocolumngrid
\bibliography{refs}

\ifarXiv
  

\section*{Supplementary material}

# Evidence that the de Almeida–Thouless transition disappears below six dimensions

Bharadwaj Vedula,[1] M. A. Moore,[2] and Auditya Sharma[1]

[1]*Department of Physics, Indian Institute of Science Education
and Research, Bhopal, Madhya Pradesh 462066, India*

[2]*Department of Physics and Astronomy, University of Manchester, Manchester M13 9PL, United Kingdom*

## I. THE SIMULATION METHOD

We now describe the technical details of our simulation process. The simulations begin with a random initial configuration, which evolves according to the method outlined in this section. To achieve rapid equilibration, we use three types of sweeps: overrelaxation (or microcanonical), heatbath, and parallel tempering. In overrelaxation and heatbath sweeps, we flip one spin at a time (single-spin-flip dynamics). To implement parallel tempering, we simulate $N_T$ copies of the system simultaneously at $N_T$ different temperatures, ranging from $T_{\min} \equiv T_1$ to $T_{\max} \equiv T_{N_T}$. To facilitate the computation of the observables discussed here, we simulate 4 sets of $N_T$ copies (2 for $h_r = 0$), labeled (1), (2), (3), and (4). Overrelaxation, heatbath, and parallel tempering sweeps are performed on all copies, with careful tracking of their labels. Every 10 overrelaxation sweeps are followed by 1 heatbath and 1 parallel tempering sweep, as overrelaxation sweeps are computationally cheaper and expedite equilibration. The simulation parameters are listed in Tables S1 and S2.

Once equilibrium is reached, we perform an equal number of sweeps in the measurement phase, making $N_{\text{sweep}}$ the total number of sweeps for the entire simulation, including both equilibration and measurement phases. The last column in the Tables S1 and S2 indicates the computer time needed to generate the data for each parameter set. The computation time mentioned here indicates the total duration needed to generate all the data using only a single core at a time (with an average clock speed of 2.6 GHz). During the measurement phase, we take one measurement for every 4 sweeps. The following sections provide detailed information on our Monte Carlo simulation procedures.

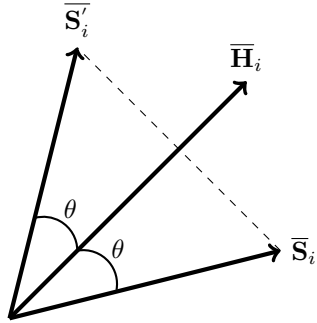

FIG. S1. The vector $\mathbf{H}_i$ is the local-field at the site $i$ which is the effective field felt by the spin $\mathbf{S}_i$ due to its interaction with all other spins. The vector $\mathbf{S}_i'$ is the reflection of $\mathbf{S}_i$ about $\mathbf{H}_i$ determined by Eq. (S2)

### A. Overrelaxation sweep

We sweep sequentially through all the lattice sites and compute the local field

$$\mathbf{H}_i = \sum_j J_{ij}\mathbf{S}_j + \mathbf{h}_i \tag{S1}$$

at each lattice site. The new spin direction $\mathbf{S}_i'$ at the $i^{\text{th}}$ lattice site is taken to be the mirror image of the vector $\mathbf{S}_i$ about $\mathbf{H}_i$, i.e.,

$$\mathbf{S}_i' = -\mathbf{S}_i + 2\frac{\mathbf{S}_i \cdot \mathbf{H}_i}{H_i^2}\mathbf{H}_i, \tag{S2}$$

where $H_i = |\mathbf{H}_i|$. Since $\mathbf{S}_i' \cdot \mathbf{H}_i = \mathbf{S}_i \cdot \mathbf{H}_i$, the energy of the system does not change due to these sweeps. Hence these sweeps are also called microcanonical sweeps. These sweeps help us in sampling out the microstates with the same energy. The process of equilibration speeds up when we include overrelaxation sweeps along with the other sweeps [1, 2].

### B. Heatbath sweep

The microcanonical sweep samples out microstates from the state-space with same energy. But, to equilibrate the system, we have to sample out the states $\{\mathbf{S}_i\}$ with different energies according to the correct Boltzmann weight $\exp(-\beta\mathcal{H}(\{\mathbf{S}_i\}))/Z$, where $\beta = 1/T$ is the inverse temperature and $Z$ is the partition function. Therefore, we perform a heatbath sweep for every 10 microcanoical sweeps [1, 2]. Similar to the microcanonical case, we sweep sequentially through the lattice and compute the local field vector $\mathbf{H}_i$ given by Eq. (S1). We can define a coordinate system with $\mathbf{H}_i$ as polar axis, $\theta$ as polar angle, and $\phi$ as azimuthal angle, such that $\phi = 0$ for the old spin vector $\mathbf{S}_i$ (see Fig. S1). The contribution of the $i^{\text{th}}$ spin to the total energy of the system is given by

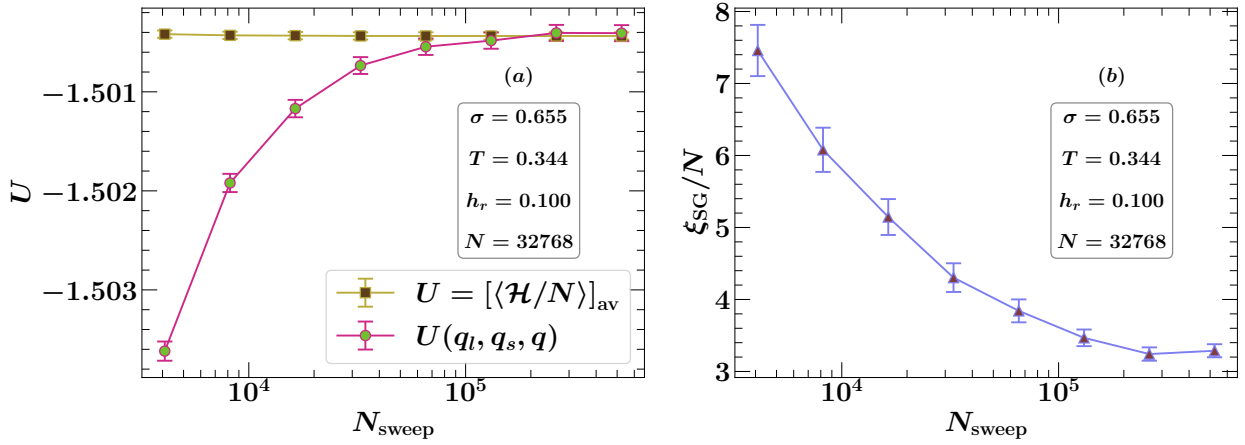

FIG. S2. The plot illustrates the test for equilibration. In (a), the energy ($U$) is plotted as a function of Monte Carlo sweeps. The brown curve, labeled $U = [\langle \mathcal{H}/N \rangle]_{\mathrm{av}}$, is derived from Eq. (S12), while the pink curve represents $U(q_l, q_s, q)$ calculated from the right-hand side of Eq. (S11). As evident from the plot, both quantities converge to the same value after $10^5$ sweeps. Once this common point is reached, the curves flatten and remain constant over time, indicating that the system has achieved thermal equilibrium. To reinforce our test, we also analyzed $\xi_{\mathrm{SG}}/N$ as a function of $N_{\mathrm{sweep}}$, shown in (b). This figure demonstrates that the $\xi_{\mathrm{SG}}$ data becomes constant at the $N_{\mathrm{sweep}}$ value where the curves in (a) converge. The data presented here are for a specific set of parameters. We repeated this test for all parameter sets listed in Tables S1 and S2, and proceeded with measurements only after confirming the system had properly equilibrated.

$E_i = -\mathbf{H}_i \cdot \mathbf{S}_i = -H_i S_i \cos \theta$ (since we are considering spins of unit length $S_i = |\mathbf{S}_i| = 1$). For the new spin direction $\mathbf{S}'_i(\theta, \phi)$, since the energy does not depend on $\phi$, we pick it uniformly from the interval $(0, 2\pi)$, and sample out $\theta$ from the probability distribution

$$f_X(x = \cos \theta) = \frac{\mathrm{e}^{-\beta E_i}}{Z} = \frac{\beta H_i S_i}{2 \sinh \beta H_i S_i} \mathrm{e}^{\beta H_i S_i x}. \tag{S3}$$

The simplest way to do this is to equate the cumulative distributive function (CDF) of $x = \cos \theta$, $F_X(x)$, to that of a uniform distribution:

$$F_X(x) = \int\limits_{-1}^{x} f_X(x') \, dx' = \Pi(r_2) = \int\limits_{0}^{r_2} \mathrm{d}r = r_2, \tag{S4}$$

where $r_2$ is a random variable sampled from a uniform distribution in the interval $(0, 1)$. Upon simplifying Eq. (S4), we get

$$x = \cos \theta = \frac{1}{\beta H_i S_i} \ln \left[ 1 + r_2 (\mathrm{e}^{2\beta H_i S_i} - 1) \right] - 1. \tag{S5}$$

To find the components of the spin vector $\mathbf{S}'_i$ in the original Cartesian coordinates, we perform a rotation about Y-axis by $\theta_H$ in the anti-clockwise direction, and then about Z-axis by $\phi_H$ in the

clockwise direction, where $\theta_H$ and $\phi_H$ are the polar and azimuthal angles of $\mathbf{H}_i$ relative to the Cartesian reference frame, i.e.,

$$\begin{pmatrix} S'_x \\ S'_y \\ S'_z \end{pmatrix} = R_Z(-\phi_H)R_Y(\theta_H) \begin{pmatrix} \sin\theta\cos\phi \\ \sin\theta\sin\phi \\ \cos\theta \end{pmatrix},$$ (S6)

where

$$R_Y(\theta_H) = \begin{pmatrix} \cos\theta_H & 0 & \sin\theta_H \\ 0 & 1 & 0 \\ -\sin\theta_H & 0 & \cos\theta_H \end{pmatrix},$$ (S7)

$$R_Z(-\phi_H) = \begin{pmatrix} \cos\phi_H & -\sin\phi_H & 0 \\ \sin\phi_H & \cos\phi_H & 0 \\ 0 & 0 & 1 \end{pmatrix}.$$ (S8)

The acceptance probability for both heatbath sweeps and microcanonical sweeps is unity, ensuring that no moves are wasted.

### C. Parallel tempering sweep

Spin glasses exhibit a complex free energy landscape, causing them to become trapped in metastable valleys at low temperatures, making true equilibration very time-consuming. At higher temperatures, thermal fluctuations allow the system to escape these valleys easily, resulting in quicker equilibration. To achieve equilibrium with the fewest moves, we perform one parallel tempering sweep for every 10 overrelaxation sweeps [1, 3]. To benefit from the parallel tempering algorithm [4, 5], we simultaneously simulate $N_T$ copies of the system at $N_T$ different temperatures $T_1 < T_2 < T_3 < \cdots < T_{N_T}$. The minimum temperature $T_1$ is the low temperature of interest for studying system behavior, while the maximum temperature $T_{N_T}$ is sufficiently high for rapid equilibration. Overrelaxation and heatbath sweeps are performed separately on each of the $N_T$ copies of the system. In the parallel tempering sweep, we compare the energies of two spin configurations at adjacent temperatures, $T_i$ and $T_{i+1}$, starting from the smallest temperature $T_1$. We swap these two spin configurations such that the detailed balance condition is satisfied. The Metropolis

probability for such a swap is

$$P(T \text{ swap}) = \min\{1, \exp(\Delta\beta\Delta E)\} \tag{S9}$$

$$= \begin{cases} \exp(\Delta\beta\Delta E), & (\text{if } \Delta\beta\Delta E < 0), \\ 1, & (\text{otherwise}), \end{cases} \tag{S10}$$

where $\Delta\beta = 1/T_i - 1/T_{i+1}$ and $\Delta E = E_i(T_i) - E_{i+1}(T_{i+1})$. In this way, a given set of spins performs a random walk in temperature space.

### D.   Checks for equilibration

In order to check whether the system has reached equilibrium, we have used a convenient test [6] which is possible because of the Gaussian nature of the interactions and the onsite external magnetic field. The relation

$$U = \frac{zJ^2}{2T}(q_l - q_s) + \frac{h_r^2}{T}\left(q - |\mathbf{S}|^2\right), \tag{S11}$$

is valid in equilibrium. Here

$$\begin{aligned} U &= \frac{1}{N}\left[\langle\mathcal{H}\rangle\right]_{\text{av}} \\ &= -\frac{1}{N}\left[\sum_{\langle i,j\rangle}\epsilon_{ij}J_{ij}\langle\mathbf{S}_i\cdot\mathbf{S}_j\rangle + \sum_{i,\mu}h_i^\mu\langle S_i^\mu\rangle\right]_{\text{av}} \end{aligned} \tag{S12}$$

is the average energy per spin,

$$q = \frac{1}{N}\sum_i\left[\langle\mathbf{S}_i\rangle\cdot\langle\mathbf{S}_i\rangle\right]_{\text{av}}$$

is the Edwards-Anderson order parameter,

$$q_l = \frac{1}{N_b}\sum_{\langle i,j\rangle}\left[\epsilon_{ij}\langle\mathbf{S}_i\cdot\mathbf{S}_j\rangle^2\right]_{\text{av}}$$

is the "link overlap", and

$$q_s = \frac{1}{N_b}\sum_{\langle i,j\rangle}\left[\epsilon_{ij}\left\langle(\mathbf{S}_i\cdot\mathbf{S}_j)^2\right\rangle\right]_{\text{av}}$$

is the "spin overlap", where $N_b = Nz/2$, and $\epsilon_{ij} = 1$ if the $i^{\text{th}}$ and $j^{\text{th}}$ spins are interacting and is zero otherwise. As the system reaches equilibrium, the two sides of Eq. (S11) approach their common equilibrium value from opposite directions.

In simulations, we assess both sides of Eq.(S11) for various numbers of Monte Carlo sweeps (MCSs), which increase exponentially, with each value being double the previous one. Averaging is done over the last half of the sweeps. Starting with a random spin configuration, the LHS of Eq.(S11) is initially small, and the RHS is very large. As the system nears equilibrium, these values converge from opposite directions, as illustrated in Fig. S2(a). We declare the system to have reached equilibrium when the averaged quantities consistently satisfy Eq. (S11) within error bars for at least the last two points. Once equilibrium is achieved, we perform an equal number of sweeps in the measurement phase, where we evaluate various quantities to study potential phase transitions in the system. To further strengthen our equilibration check, we analysed $\xi_{SG}$ as a function of the Monte Carlo sweeps. Our data in Fig. S2(b) show that the $\xi_{SG}$ curve flattens off for the same number of sweeps where the two sides of Eq. (S11) start to agree.

## II. PARAMETERS OF THE SIMULATIONS

The simulation parameters for our study are summarized in Tables S1 and S2, detailing the distinct setups for temperature-based and field-based runs. Table S1 lists the parameters used in simulations over a temperature range, with $N_{samp}$ indicating the number of disorder samples and $N_{sweep}$ representing the number of over-relaxation Monte Carlo sweeps per sample. The system equilibrates during the initial half of the sweeps, while measurements are gathered in the latter half, sampled every fourth sweep. Additionally, the temperature range is specified by $T_{min}$ and $T_{max}$, with $N_T$ denoting the number of temperatures applied in parallel tempering. The total CPU time $t_{tot}$ required for data generation at each system size is also noted. Table S2 presents the parameters for simulations conducted at fixed temperatures while varying the field $h_r$. Here, $N(h_r)$ reflects the count of field values within the specified range $(h_r(\min), h_r(\max))$. The equilibration times and number of disorder samples vary across field values, with their ranges defined as $N_{sweep}(\min,\max)$ and $N_{samp}(\min,\max)$, respectively. The total CPU time for data collection, $t_{tot}$, is similarly provided for each system size. Together, these tables outline the computational setup and resource allocation essential for our simulation data.

TABLE S1. Parameters of the simulations. $N_{\text{samp}}$ is the number of disorder samples, $N_{\text{sweep}}$ is the number of over-relaxation Monte Carlo sweeps for a single disorder sample. The system is equilibrated over the first half of the sweeps, and measurements are done over the last half of the sweeps with a measurement performed every four over-relaxation sweeps. $T_{\text{min}}$ and $T_{\text{max}}$ are the lowest and highest temperatures simulated, and $N_T$ is the number of temperatures used for parallel tempering. $t_{\text{tot}}$ is the total CPU time consumed in hours to generate data for a particular system size.

| $\sigma$ | $h_r$ | $N$ | $N_{\text{samp}}$ | $N_{\text{sweep}}$ | $T_{\text{min}}$ | $T_{\text{max}}$ | $N_T$ | $t_{\text{tot}}$ (hrs) |
|---|---|---|---|---|---|---|---|---|
| 0.6 | 0 | 128 | 12000 | 256 | 0.2 | 0.7 | 40 | 1.2 |
| 0.6 | 0 | 256 | 12000 | 512 | 0.2 | 0.7 | 40 | 4.89 |
| 0.6 | 0 | 512 | 12000 | 1024 | 0.2 | 0.7 | 40 | 21.57 |
| 0.6 | 0 | 1024 | 12000 | 2048 | 0.2 | 0.7 | 40 | 48.58 |
| 0.6 | 0 | 2048 | 9600 | 4096 | 0.2 | 0.7 | 40 | 235.21 |
| 0.6 | 0 | 4096 | 7200 | 8192 | 0.3 | 0.7 | 40 | 1109.96 |
| 0.6 | 0 | 8192 | 3120 | 16384 | 0.3 | 0.7 | 50 | 2611.03 |
| 0.6 | 0 | 16384 | 1200 | 32768 | 0.35 | 0.7 | 55 | 6453.1 |
| 0.6 | 0 | 32768 | 408 | 65536 | 0.35 | 0.7 | 60 | 7663.65 |
| 0.63 | 0 | 128 | 8000 | 512 | 0.2 | 0.7 | 40 | 1.24 |
| 0.63 | 0 | 256 | 8000 | 1024 | 0.2 | 0.7 | 40 | 5.81 |
| 0.63 | 0 | 512 | 8000 | 2048 | 0.2 | 0.7 | 40 | 29.16 |
| 0.63 | 0 | 1024 | 8000 | 8192 | 0.2 | 0.7 | 40 | 191.87 |
| 0.63 | 0 | 2048 | 8000 | 8192 | 0.3 | 0.7 | 40 | 460.11 |
| 0.63 | 0 | 4096 | 4560 | 16384 | 0.32 | 0.68 | 40 | 1156.2 |
| 0.63 | 0 | 8192 | 3193 | 32768 | 0.36 | 0.66 | 42 | 6183.85 |
| 0.63 | 0 | 16384 | 2432 | 32768 | 0.38 | 0.66 | 44 | 9954.44 |
| 0.64 | 0 | 128 | 24000 | 512 | 0.2 | 0.7 | 40 | 3.14 |
| 0.64 | 0 | 256 | 24000 | 1024 | 0.2 | 0.7 | 40 | 17.33 |
| 0.64 | 0 | 512 | 22400 | 2048 | 0.2 | 0.7 | 40 | 70.61 |
| 0.64 | 0 | 1024 | 11200 | 8192 | 0.2 | 0.7 | 40 | 305.41 |
| 0.64 | 0 | 2048 | 16000 | 8192 | 0.3 | 0.7 | 40 | 815.66 |
| 0.64 | 0 | 4096 | 12000 | 16384 | 0.32 | 0.68 | 40 | 3751.7 |
| 0.64 | 0 | 8192 | 3360 | 32768 | 0.36 | 0.66 | 42 | 6323.48 |
| 0.64 | 0 | 16384 | 2240 | 32768 | 0.38 | 0.66 | 44 | 9724.53 |
| 0.65 | 0 | 128 | 9600 | 512 | 0.3 | 0.7 | 40 | 1.96 |
| 0.65 | 0 | 256 | 9600 | 1024 | 0.3 | 0.7 | 40 | 7.83 |
| 0.65 | 0 | 512 | 9600 | 2048 | 0.3 | 0.7 | 40 | 32.98 |
| 0.65 | 0 | 1024 | 33600 | 4096 | 0.3 | 0.7 | 44 | 457.63 |
| 0.65 | 0 | 2048 | 33600 | 8192 | 0.3 | 0.7 | 40 | 2212.92 |
| 0.65 | 0 | 4096 | 19200 | 16384 | 0.32 | 0.68 | 40 | 6242.19 |
| 0.65 | 0 | 8192 | 15054 | 16384 | 0.36 | 0.66 | 42 | 14652.7 |
| 0.65 | 0 | 16384 | 10526 | 32768 | 0.38 | 0.66 | 44 | 42378.9 |
| 0.655 | 0 | 128 | 24000 | 512 | 0.2 | 0.7 | 40 | 3.05 |
| 0.655 | 0 | 256 | 24000 | 1024 | 0.2 | 0.7 | 40 | 14.69 |
| 0.655 | 0 | 512 | 16000 | 2048 | 0.2 | 0.7 | 40 | 43.48 |
| 0.655 | 0 | 1024 | 21920 | 4096 | 0.3 | 0.7 | 40 | 291.4 |
| 0.655 | 0 | 2048 | 20400 | 8192 | 0.3 | 0.7 | 40 | 1143.21 |
| 0.655 | 0 | 4096 | 18252 | 16384 | 0.32 | 0.68 | 40 | 6818.29 |
| 0.655 | 0 | 8192 | 12260 | 16384 | 0.36 | 0.66 | 42 | 11835.9 |
| 0.655 | 0 | 16384 | 5278 | 32768 | 0.38 | 0.66 | 44 | 22993.1 |
| 0.655 | 0 | 32768 | 3646 | 32768 | 0.38 | 0.66 | 44 | 29010.7 |

TABLE S2. Parameters of the simulations done at fixed temperature $T$ and varying field $h_r$. $N(h_r)$ is the number of values of field taken in the range $h_r(\text{min,max})$. The equilibration times are different for different values of the field $h_r$, which lie in the range $N_{\text{sweep}}(\text{min,max})$. The number of disorder samples for different fields lie in the range $N_{\text{samp}}(\text{min,max})$. $t_{\text{tot}}$ is the total CPU time consumed in hours to generate data for a particular system size.

| $\sigma$ | $T$ | $N$ | $h_r(\text{min,max})$ | $N(h_r)$ | $N_{\text{sweep}}(\text{min,max})$ | $N_{\text{samp}}(\text{min,max})$ | $t_{\text{tot}}(\text{hrs})$ |
|---|---|---|---|---|---|---|---|
| 0.6 | 0.38 | 128 | (0.0100, 9.0000) | 27 | (1024, 1024) | (5000, 20000) | 2.55 |
| 0.6 | 0.38 | 256 | (0.0100, 9.0000) | 27 | (2048, 2048) | (5000, 30000) | 18.5 |
| 0.6 | 0.38 | 512 | (0.0100, 9.0000) | 27 | (2048, 2048) | (5000, 30000) | 40.78 |
| 0.6 | 0.38 | 1024 | (0.0100, 9.0000) | 30 | (4096, 4096) | (4000, 30000) | 181.68 |
| 0.6 | 0.38 | 2048 | (0.0100, 9.0000) | 31 | (4096, 8192) | (2500, 40000) | 3506.91 |
| 0.6 | 0.38 | 4096 | (0.0100, 9.0000) | 31 | (8192, 32768) | (4000, 75200) | 9948.45 |
| 0.6 | 0.38 | 8192 | (0.0100, 9.0000) | 27 | (16384, 262144) | (1200, 37050) | 88881.3 |
| 0.6 | 0.38 | 16384 | (0.0100, 9.0000) | 29 | (32768, 524288) | (240, 12749) | 151254 |
| 0.6 | 0.38 | 32768 | (0.1000, 0.1400) | 3 | (131072, 1048576) | (3813, 7596) | 155371 |
| 0.63 | 0.364 | 128 | (0.0100, 9.0000) | 28 | (512, 512) | (4000, 16000) | 2.07 |
| 0.63 | 0.364 | 256 | (0.0100, 9.0000) | 33 | (1024, 1024) | (4000, 32000) | 16.06 |
| 0.63 | 0.364 | 512 | (0.0100, 9.0000) | 34 | (2048, 2048) | (4000, 32000) | 52.31 |
| 0.63 | 0.364 | 1024 | (0.0100, 9.0000) | 33 | (4096, 4096) | (5000, 32000) | 200.35 |
| 0.63 | 0.364 | 2048 | (0.0100, 9.0000) | 33 | (4096, 8192) | (4000, 36000) | 967.71 |
| 0.63 | 0.364 | 4096 | (0.0100, 9.0000) | 32 | (8192, 32768) | (4000, 37440) | 7357.41 |
| 0.63 | 0.364 | 8192 | (0.0100, 9.0000) | 33 | (16384, 131072) | (643, 34251) | 56078.5 |
| 0.63 | 0.364 | 16384 | (0.0100, 9.0000) | 30 | (32768, 524288) | (640, 18941) | 94942.1 |
| 0.63 | 0.364 | 32768 | (0.0900, 0.1200) | 4 | (262144, 524288) | (4892, 14287) | 159218 |
| 0.64 | 0.357 | 128 | (0.0100, 9.0000) | 27 | (1024, 1024) | (5000, 40000) | 4.56 |
| 0.64 | 0.357 | 256 | (0.0100, 9.0000) | 27 | (2048, 2048) | (2000, 40000) | 19.01 |
| 0.64 | 0.357 | 512 | (0.0400, 9.0000) | 25 | (2048, 2048) | (2000, 50000) | 48.42 |
| 0.64 | 0.357 | 1024 | (0.0100, 9.0000) | 28 | (4096, 4096) | (2000, 40000) | 190.31 |
| 0.64 | 0.357 | 2048 | (0.0100, 9.0000) | 28 | (4096, 8192) | (1000, 40000) | 791.55 |
| 0.64 | 0.357 | 4096 | (0.0400, 9.0000) | 25 | (8192, 16384) | (1000, 36000) | 4250.85 |
| 0.64 | 0.357 | 8192 | (0.0100, 9.0000) | 28 | (16384, 131072) | (1600, 37120) | 23608.4 |
| 0.64 | 0.357 | 16384 | (0.0100, 9.0000) | 28 | (32768, 262144) | (320, 12800) | 58060.3 |
| 0.65 | 0.35 | 128 | (0.0100, 9.0000) | 29 | (1024, 1024) | (4000, 12000) | 10.12 |
| 0.65 | 0.35 | 256 | (0.0100, 9.0000) | 29 | (2048, 2048) | (4000, 12000) | 14.64 |
| 0.65 | 0.35 | 512 | (0.0100, 9.0000) | 29 | (4096, 4096) | (4000, 12000) | 58.13 |
| 0.65 | 0.35 | 1024 | (0.0100, 0.3000) | 14 | (8192, 8192) | (4000, 10000) | 139.38 |
| 0.65 | 0.35 | 2048 | (0.0100, 0.3000) | 14 | (16384, 16384) | (4800, 24000) | 800.5 |
| 0.65 | 0.35 | 4096 | (0.0100, 9.0000) | 29 | (32768, 32768) | (1200, 24000) | 4786.42 |
| 0.65 | 0.35 | 8192 | (0.0100, 0.3000) | 14 | (65536, 131072) | (800, 21551) | 27908.2 |
| 0.65 | 0.35 | 16384 | (0.0100, 0.3000) | 14 | (262144, 262144) | (551, 11602) | 74407.7 |
| 0.655 | 0.344 | 128 | (0.0100, 9.0000) | 27 | (1024, 1024) | (4000, 44000) | 6.03 |
| 0.655 | 0.344 | 256 | (0.0100, 9.0000) | 29 | (2048, 2048) | (4000, 44000) | 30.18 |
| 0.655 | 0.344 | 512 | (0.0100, 9.0000) | 29 | (4096, 4096) | (2000, 60000) | 80.09 |
| 0.655 | 0.344 | 1024 | (0.0100, 9.0000) | 30 | (8192, 8192) | (2000, 48000) | 265.81 |
| 0.655 | 0.344 | 2048 | (0.0100, 9.0000) | 30 | (16384, 16384) | (1000, 11600) | 1838.4 |
| 0.655 | 0.344 | 4096 | (0.0100, 9.0000) | 29 | (16384, 32768) | (960, 36960) | 7440.49 |
| 0.655 | 0.344 | 8192 | (0.0100, 9.0000) | 32 | (16384, 65536) | (480, 42800) | 42173.3 |
| 0.655 | 0.344 | 16384 | (0.0100, 9.0000) | 32 | (32768, 262144) | (576, 20479) | 191977 |
| 0.655 | 0.344 | 32768 | (0.0100, 9.0000) | 29 | (65536, 524288) | (256, 14707) | 272176 |
| 0.655 | 0.344 | 65536 | (0.0100, 0.2000) | 8 | (524288, 1048576) | (192, 3845) | 245877 |

## III.  $\xi_{\mathrm{SG}}$ DATA

We believe that the results for $h_{\mathrm{AT}}$ derived from $\xi_{\mathrm{SG}}$ data are badly affected by finite size effects and are thus less reliable than the values obtained from $\chi_{\mathrm{SG}}$, possibly for the reasons given at the end of Sec. III of the main text. That is why these analyses have been relegated to this Appendix.

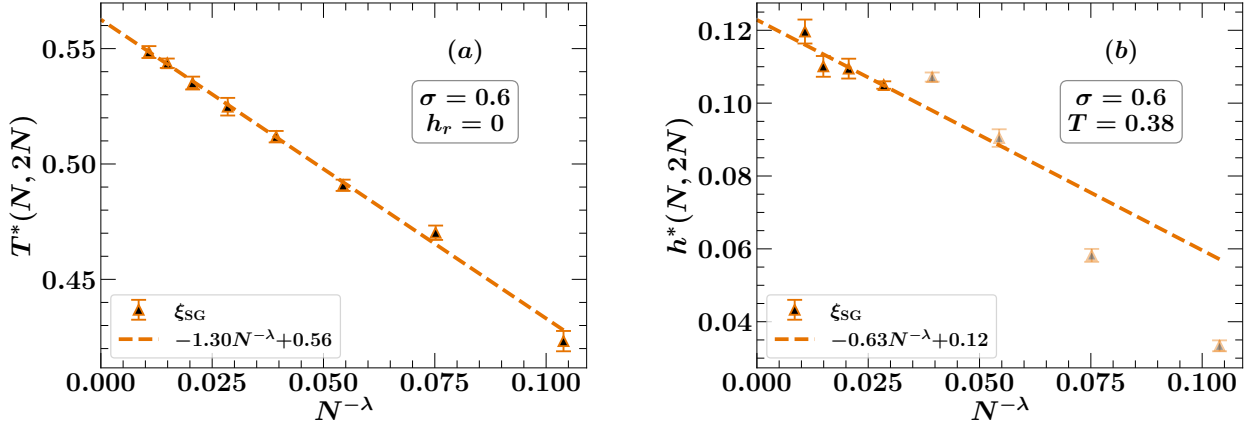

FIG. S3. Plots showing (a) $T^*(N, 2N)$ and (b) $h^*(N, 2N)$ data obtained from $\xi_{\mathrm{SG}}$ as a function of $N^{-\lambda}$ for $\sigma = 0.600$. The value of $\lambda = 0.467$ used here is the same as the value used for the corresponding $\chi_{\mathrm{SG}}$ data in Figs. 2(b) and 3(b). Both the data sets are fitted with a straight line and the resulting values are $T_c = 0.5626 \pm 0.0017$ from (a) and $h_{\mathrm{AT}}(T = 0.380) = 0.1229 \pm 0.0038$ from (b). The blurred points in (b) are excluded from the linear fitting.

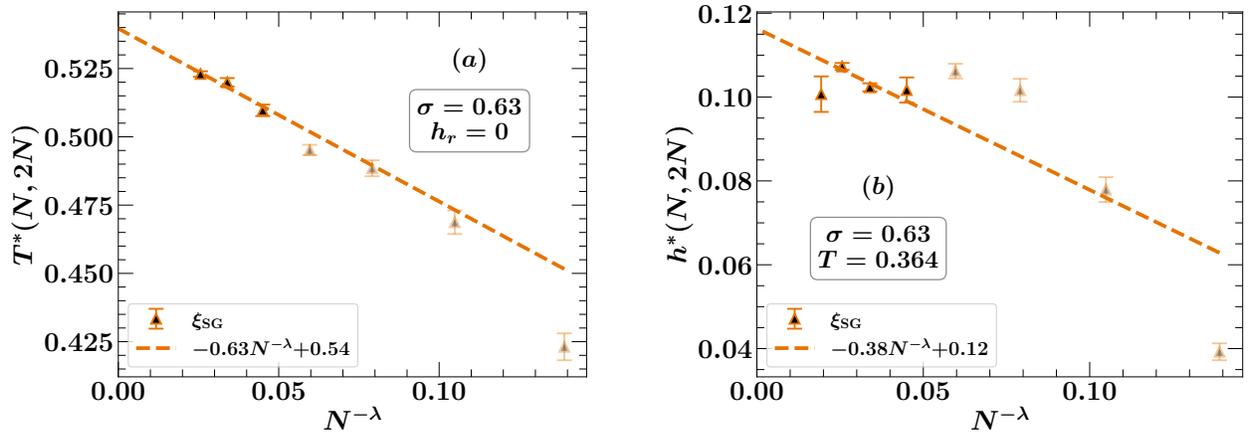

FIG. S4. Plots showing (a) $T^*(N, 2N)$ and (b) $h^*(N, 2N)$ data obtained from $\xi_{\mathrm{SG}}$ as a function of $N^{-\lambda}$ for $\sigma = 0.630$. The value of $\lambda = 0.407$ used here is the same as the value used for the corresponding $\chi_{\mathrm{SG}}$ data in Figs. 4(b) and 5(b). Both the data sets are fitted with a straight line and the resulting values are $T_c = 0.5396 \pm 0.0035$ from (a) and $h_{\mathrm{AT}}(T = 0.364) = 0.1164 \pm 0.0035$ from (b). The blurred points in (a) and (b) are excluded from the linear fitting.

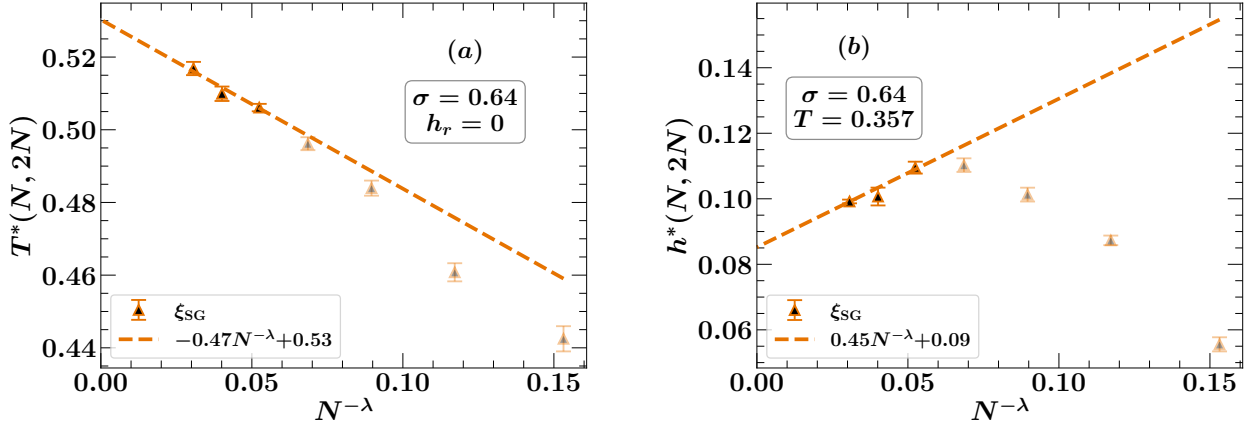

FIG. S5. Plots showing (a) $T^*(N, 2N)$ and (b) $h^*(N, 2N)$ data obtained from $\xi_{\mathrm{SG}}$ as a function of $N^{-\lambda}$ for $\sigma = 0.640$. The value of $\lambda = 0.387$ used here is the same as the value used for the corresponding $\chi_{\mathrm{SG}}$ data in Figs. 6(b) and 7(b). Both the data sets are fitted with a straight line and the resulting values are $T_c = 0.5303 \pm 0.0043$ from (a) and $h_{\mathrm{AT}}(T = 0.357) = 0.0852 \pm 0.0028$ from (b). The blurred points in (a) and (b) are excluded from the linear fitting.

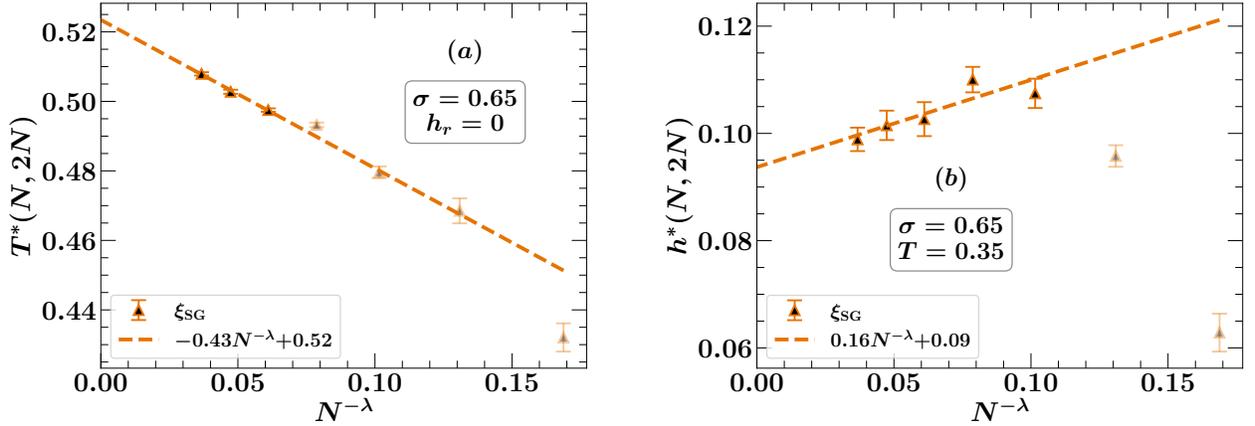

FIG. S6. Plots showing (a) $T^*(N, 2N)$ and (b) $h^*(N, 2N)$ data obtained from $\xi_{\mathrm{SG}}$ as a function of $N^{-\lambda}$ for $\sigma = 0.650$. The value of $\lambda = 0.367$ used here is the same as the value used for the corresponding $\chi_{\mathrm{SG}}$ data in Figs. 8(b) and 9(b). Both the data sets are fitted with a straight line and the resulting values are $T_c = 0.5235 \pm 0.0014$ from (a) and $h_{\mathrm{AT}}(T = 0.350) = 0.0937 \pm 0.0033$ from (b). The blurred points in (a) and (b) are excluded from the linear fitting.

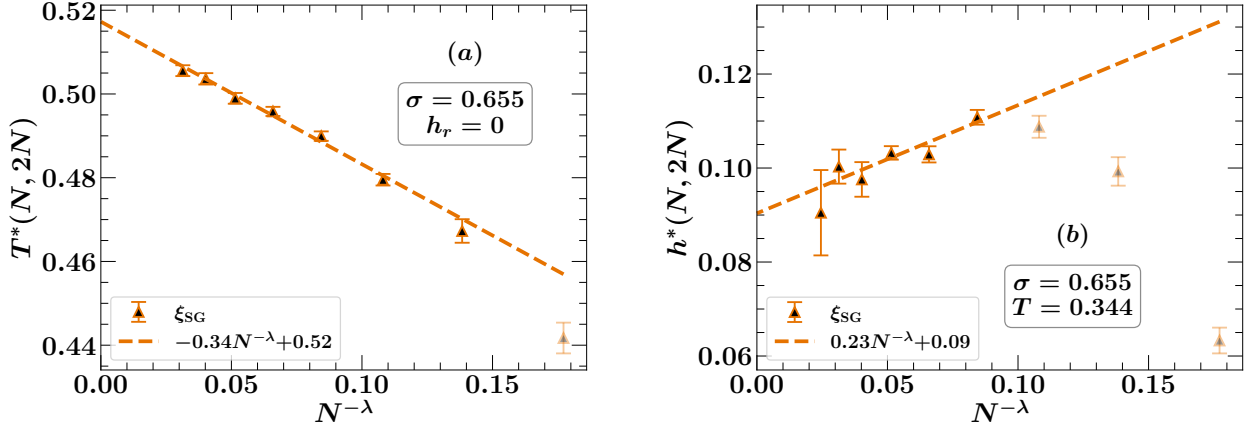

FIG. S7. Plots showing (a) $T^*(N, 2N)$ and (b) $h^*(N, 2N)$ data obtained from $\xi_{\text{SG}}$ as a function of $N^{-\lambda}$ for $\sigma = 0.655$. The value of $\lambda = 0.357$ used here is the same as the value used for the corresponding $\chi_{\text{SG}}$ data in Figs. 10(b) and 11(b). Both the data sets are fitted with a straight line and the resulting values are $T_c = 0.5173 \pm 0.0013$ from (a) and $h_{\text{AT}}(T = 0.344) = 0.0904 \pm 0.0033$ from (b). The blurred points in (a) and (b) are excluded from the linear fitting.

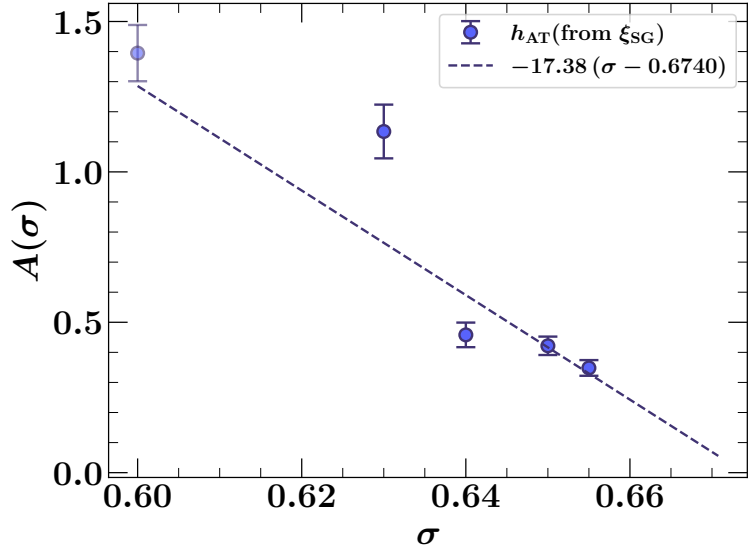

FIG. S8. The plot shows $A(\sigma)$ versus $\sigma$, similar to Fig. 12 in the main text, but here the $A(\sigma)$ data is derived from the $\xi_{\text{SG}}$ dataset (see, for example, Fig. S4). The $A(\sigma)$ data, excluding $\sigma = 0.600$, is fitted with a straight line. The slope and $y$-intercept of the fit are $m = -17.381 \pm 2.614$ and $c = 11.714 \pm 1.698$, respectively. The correlation coefficient between $m$ and $c$ is $\rho_{mc} \approx -1$, signifying a strong negative correlation between the slope and intercept. The fit yields a reduced chi-squared value of $\chi^2/N_{\text{dof}} = 14.087$, indicating a poor fit. This is the reason this result has been moved to the appendix. The line intersects the $\sigma$-axis at $\sigma_l = 0.674 \pm 0.004$, corresponding to $d_l = 5.748 \pm 0.125$. In performing the linear fits, the data point at $\sigma = 0.600$ is excluded (and blurred), as it lies outside the linear region, which is valid only for $\sigma$ values near $2/3$. Detailed calculations for the uncertainties in $\sigma_l$ and $d_l$ are provided in Sec. IV.

## IV. PROPAGATION OF UNCERTAINTY

In this section, we provide details on how to compute the uncertainties associated with the estimates of $A(\sigma)$, $\sigma_l$, and $d_l$. We begin by outlining the general procedure for propagating uncertainties in functions of two variables [7, 8].

Let $f(a, b)$ be a differentiable function of two variables $a$ and $b$. The uncertainty in $f$, denoted by $\delta f$, can be calculated using the expression:

$$\delta f \approx \sqrt{\left(\frac{\partial f}{\partial a}\right)^2 (\delta a)^2 + \left(\frac{\partial f}{\partial b}\right)^2 (\delta b)^2 + 2\frac{\partial f}{\partial a}\frac{\partial f}{\partial b}\delta_{ab}}, \tag{S13}$$

where $\delta a$ and $\delta b$ represent the uncertainties (standard errors) associated with $a$ and $b$, respectively, and $\delta_{ab}$ is the covariance between $a$ and $b$. The covariance term $\delta_{ab}$ accounts for the correlation between $a$ and $b$. When $\delta_{ab} > 0$, the variables $a$ and $b$ are positively correlated, meaning they tend to increase or decrease together. Conversely, when $\delta_{ab} < 0$, $a$ and $b$ are negatively correlated, so an increase in $a$ typically corresponds to a decrease in $b$, and vice versa.

The strength and direction of the correlation between $a$ and $b$ can be quantified by the correlation coefficient $\rho_{ab}$, which is defined as:

$$\rho_{ab} = \frac{\delta_{ab}}{\delta a\, \delta b}. \tag{S14}$$

The correlation coefficient $\rho_{ab}$ lies within the interval $[-1, 1]$. A value of $\rho_{ab} = 1$ indicates a perfect positive correlation, while $\rho_{ab} = -1$ indicates a perfect negative correlation. When $\rho_{ab} = 0$, the variables $a$ and $b$ are uncorrelated or independent, and the covariance term $\delta_{ab}$ vanishes. In such cases, Eq. (S13) simplifies to:

$$\delta f \approx \sqrt{\left(\frac{\partial f}{\partial a}\right)^2 (\delta a)^2 + \left(\frac{\partial f}{\partial b}\right)^2 (\delta b)^2}. \tag{S15}$$

These formulas form the basis for computing uncertainties in derived quantities. In the following subsections, we use this approach to derive the specific uncertainties for $A(\sigma)$, $\sigma_l$, and $d_l$.

### A. Uncertainty in $A(\sigma)$

In this subsection, we derive the uncertainty associated with the quantity $A(\sigma)$. As shown in Eq. (1) of the main text, $A(\sigma)$ is given by:

$$A(\sigma) = \frac{\left(\dfrac{h_{\mathrm{AT}}}{T_c}\right)^2}{\left(1 - \dfrac{T}{T_c}\right)^{\zeta}}, \tag{S16}$$

where $h_{\text{AT}}$ is the transition field, $T_c$ is the zero-field transition temperature, and $\zeta$ is an exponent defined as:

$$\zeta = \begin{cases} 3, & \sigma < 5/8, \\ \dfrac{d_{\text{eff}}}{2} - 1, & 5/8 < \sigma < 2/3. \end{cases}$$

For each value of $\sigma$, we first compute the zero-field transition temperature $T_c$, and then determine the transition field $h_{\text{AT}}(T)$ by fixing the temperature at $T = 0.67\, T_c$. Since $T_c$ and $h_{\text{AT}}$ are measured independently, they are uncorrelated. Therefore, using the general formula for uncertainty propagation in Eq. (S15), the uncertainty in $A(\sigma)$, denoted as $\delta A$, is given by:

$$\delta A = \sqrt{\left(\frac{\partial A}{\partial h_{\text{AT}}}\right)^2 (\delta h_{\text{AT}})^2 + \left(\frac{\partial A}{\partial T_c}\right)^2 (\delta T_c)^2}, \tag{S17}$$

where $\delta h_{\text{AT}}$ and $\delta T_c$ represent the uncertainties in $h_{\text{AT}}$ and $T_c$, respectively.

To compute $\delta A$, we need the partial derivatives of $A(\sigma)$ with respect to $h_{\text{AT}}$ and $T_c$. These derivatives are given by:

$$\frac{\partial A}{\partial h_{\text{AT}}} = \frac{2A}{h_{\text{AT}}}, \quad \frac{\partial A}{\partial T_c} = -\frac{A}{T_c}\left(2 + \frac{\zeta T}{T_c - T}\right). \tag{S18}$$

By substituting these derivatives into Eq. (S17), we can compute the total uncertainty $\delta A$ for any given values of $h_{\text{AT}}$, $T_c$, and their respective uncertainties.

### B. Uncertainty in $\sigma_l$ and $d_l$

In this subsection, we derive the uncertainties associated with $\sigma_l$ and $d_l$, which arise from the linear fit of $A(\sigma)$ versus $\sigma$. When fitting the data with a straight line, we obtain the slope $m$, the $y$-intercept $c$, and their associated uncertainties, $\delta m$ and $\delta c$, respectively. Additionally, the covariance between $m$ and $c$, denoted as $\delta_{mc}$, is also obtained from the linear fit and can be expressed in terms of the correlation coefficient $\rho_{mc}$.

The quantity $\sigma_l$ represents the point where the fitted line intersects the $\sigma$-axis. Therefore, $\sigma_l$ is given by:

$$\sigma_l = -\frac{c}{m}. \tag{S19}$$

To propagate the uncertainties in $m$ and $c$ to $\sigma_l$, we use the general formula for uncertainty propagation. The partial derivatives of $\sigma_l$ with respect to $m$ and $c$ are:

$$\frac{\partial \sigma_l}{\partial m} = \frac{c}{m^2}, \quad \frac{\partial \sigma_l}{\partial c} = -\frac{1}{m}. \tag{S20}$$

Substituting these partial derivatives into the uncertainty propagation formula given by Eq. (S13), the uncertainty in $\sigma_l$, denoted by $\delta\sigma_l$, is:

$$\delta\sigma_l = \sqrt{\left(\frac{c}{m^2}\delta m\right)^2 + \left(\frac{\delta c}{m}\right)^2 - \frac{2c}{m^3}\delta_{mc}}. \tag{S21}$$

In practice, for both the line fits shown in Fig. 12 of the main text and Fig. S8, the correlation coefficient $\rho_{mc}$ is approximately $-1$. This strong negative correlation implies, according to Eq. (S14), that the covariance $\delta_{mc}$ is given by:

$$\delta_{mc} \approx -\delta m\delta c. \tag{S22}$$

Substituting this into Eq. (S21), the expression for the uncertainty $\delta\sigma_l$ simplifies to:

$$\begin{aligned}
\delta\sigma_l &\approx \sqrt{\left(\frac{c}{m^2}\delta m\right)^2 + \left(\frac{\delta c}{m}\right)^2 + \frac{2c}{m^3}\delta m\delta c}, \\
&= \sqrt{\left(\frac{c}{m^2}\delta m + \frac{1}{m}\delta c\right)^2}, \\
&= \left|\frac{c}{m^2}\delta m + \frac{1}{m}\delta c\right|. 
\end{aligned} \tag{S23}$$

The effective spatial dimension $d_l$ corresponding to $\sigma_l$ is defined as:

$$d_l = \frac{2}{2\sigma_l - 1}. \tag{S24}$$

To compute the uncertainty in $d_l$, denoted as $\delta d_l$, we use the propagation of uncertainties formula. Since $d_l$ depends solely on $\sigma_l$, the uncertainty $\delta d_l$ is given by:

$$\delta d_l = \left|\frac{\partial d_l}{\partial \sigma_l}\right|\delta\sigma_l. \tag{S25}$$

The partial derivative of $d_l$ with respect to $\sigma_l$ is:

$$\frac{\partial d_l}{\partial \sigma_l} = -\frac{4}{(2\sigma_l - 1)^2}. \tag{S26}$$

Substituting this into Eq. (S25), the expression for $\delta d_l$ becomes:

$$\delta d_l = \frac{4}{(2\sigma_l - 1)^2}\,\delta\sigma_l = d_l^2\,\delta\sigma_l. \tag{S27}$$

In summary, the uncertainties $\delta\sigma_l$ and $\delta d_l$ are determined by the uncertainties in the slope $m$ and intercept $c$ from the linear fit of $A(\sigma)$ versus $\sigma$.


\fi

\end{document}